\documentclass[traditabstract]{aa}
\usepackage{graphicx}
\usepackage{subfloat}
\usepackage{graphicx}
\usepackage{caption}
\usepackage{subcaption}
\usepackage{amsmath}
\usepackage{natbib}
\usepackage{color}
\usepackage[english]{babel}
\bibpunct{(}{)}{;}{a}{}{,} 

\begin{document}

\title{When the Milky Way turned off the lights: APOGEE provides evidence of star formation quenching in our Galaxy}

\titlerunning{When the Milky Way turned-off the lights}

\author{M. Haywood\inst{1},  M.~D. Lehnert\inst{2},  P. Di Matteo\inst{1}, O. Snaith\inst{3}, M. Schultheis\inst{4}, D. Katz\inst{1}, A. G\'omez\inst{1}}

\authorrunning{Haywood et al.}

\institute{GEPI, Observatoire de Paris, CNRS, Universit$\rm \acute{e}$  Paris Diderot, 5 place Jules Janssen, 92190 Meudon, France\\
\email{Misha.Haywood@obspm.fr}
\and
Institut d'Astrophysique de Paris, CNRS UMR 7095, Universit$\rm \acute{e}$
Pierre et Marie Curie, 98 bis Bd Arago, 75014 Paris, France
\and
School of Physics, Korea Institute for Advanced Study, 85 Hoegiro, 
Dongdaemun-gu, Seoul 02455, Republic of Korea  
\and
Universit\'e de Nice Sophia-Antipolis, CNRS, Observatoire de C\^ote d’Azur, Laboratoire Lagrange, 06304 Nice Cedex 4, France}

\date{Accepted, Received}

\abstract{
Quenching, the cessation of star formation, is one of the most significant
events in the life cycle of galaxies. While quenching is generally
thought to be linked to their central regions, the mechanism responsible
for it is not known and may not even be unique.  We show here the first
evidence that the Milky Way experienced a generalised quenching of its
star formation at the end of its thick-disk formation $\sim$9 Gyr ago. The
fossil record imprinted on the elemental abundances of stars studied in
the solar vicinity and as part of the APOGEE survey (APOGEE is part of the Sloan Digital Sky Survey III) 
reveals indeed that
in less than $\sim$2 Gyr (from 10 to 8 Gyr ago) the star formation rate
in our Galaxy dropped by an order of magnitude.  Because of the tight
correlation that exists between age and $\alpha$ abundance, the general
cessation of the star formation activity reflects in the dearth of
stars along the inner-disk sequence in the [Fe/H]-[$\alpha$/Fe] plane.
Before this phase, which lasted about 1.5 Gyr, the Milky Way was
actively forming stars. Afterwards, the star formation resumed at a much
lower level to form the thin disk.  These events observed in our Galaxy
are very well matched by the latest observation of MW-type progenitors
at high redshifts.  In late-type galaxies, the quenching mechanism is
believed to be related to a long and secular exhaustion of gas. Our
results show that in the Milky Way, the shut-down occurred on a much
shorter timescale, while the chemical continuity between the stellar
populations formed before and after the quenching indicates that it is
not the exhaustion of the gas that was responsible for the cessation
of the star formation.  While quenching is generally associated with
spheroids in the literature, our results show that it also occurs in
galaxies  like the Milky Way, where the classical bulge is thought to be small or non-existent, possibly when they
are undergoing a morphological transition from thick to thin disks.
Given the demographics of late-type galaxies in the local Universe, in
which classical bulges are rare, we suggest further that this may hold
true generally in galaxies with mass lower than or approximately
of $M^*$,
where quenching could directly be a consequence of thick-disk formation,
while quenching may be related to development of spheroids in higher mass
galaxies.  We emphasize that the quenching phase in the Milky Way could
be contemporaneous with, and related to, the formation of the bar, at
the end of the thick-disk phase.  We sketch a scenario on how a strong
bar may inhibit star formation.
}

\keywords{Galaxy: abundances --- Galaxy: disk --- Galaxy: evolution --- galaxies: evolution}
\maketitle

\section{Introduction}

Galaxies are observed to quench their star formation activity at an
epoch that depends on their mass \citep[e.g][]{cimatti06}, with the
most massive galaxies quenching at z$\sim$2-3 \citep{cimatti04}.
Whether this quenching is a complete shut-down of star formation or a
more gradual transition from an active phase of star formation to a more
quiescent one is still debated.  Quenching in massive galaxies
seems to be rapid \cite[e.g][]{mancini15}, and possibly slower in
spiral galaxies \citep{schawinski14}.  
Various mechanisms have been proposed to explain why
galaxies quench as a function of mass and epoch. The first class of proposed 
mechanisms either removes the gas from the system or prevent further gas accretion.
This is the case for powerful feedback from
active galactic nuclei, the so-called AGN feedback \citep{granato04}, which could
sweep the gas from galaxies, but it is thought 
to be effective only for the most massive galaxies \citep[see e.g][] {forster14, gabor14},
while feedback from supernovae could be effective for less massive objects \citep[][]{kaviraj07}.
Gas accretion can be halted by heating 
circum-galactic gas (e.g. radio-mode AGN feedback; Croton et al. 2006)
or prevent accretion of additional cold gas through virial shock
heating \citep[][]{birnboim03}. The second class of mechanisms inhibits star formation 
either through stabilization of a disk galaxy as it is transformed to a spheroid
(also known as morphological quenching; Martig et al. 2009) or dust heating by
radiation of low-mass stars \citep[][]{kajisawa15,conroy15}, 
or high levels of turbulence generated by an AGN \citep[][]{guillard15}.

Recent results also
suggest that quenching in disks is related to the mass build-up of bulges and
morphological change in galaxies, perhaps marking the end of the bulge
building phase \citep{bluck14,lang14,abramson14,bell12}. 
For Milky Way-type galaxies, selected to have co-moving number densities
similar to that of galaxies with stellar mass similar to the Milky Way (hereafter MW)
at z$\sim$0, quenching seems to occur at z$\sim$1.6 \citep{morishita15} in the inner
parts of disks ($<$2.5 kpc).
Because most of the star formation since z$\sim$2-3 occurred in Milky
Way-like disk galaxies \citep[e.g][]{nelson13}, which are the dominant
type on the main sequence, understanding if and how star formation
quenching occurred in our Galaxy is of paramount importance.
We argue in the present paper that the MW provides the nearest
example of quenching in a galaxy, occurring 9-10 Gyr ago at 
the transition from the thick to the thin disks, in agreement with 
what is measured on high-redshift galaxies.

\begin{figure}
\includegraphics[trim=30 40 0 10,clip,width=8.5cm]{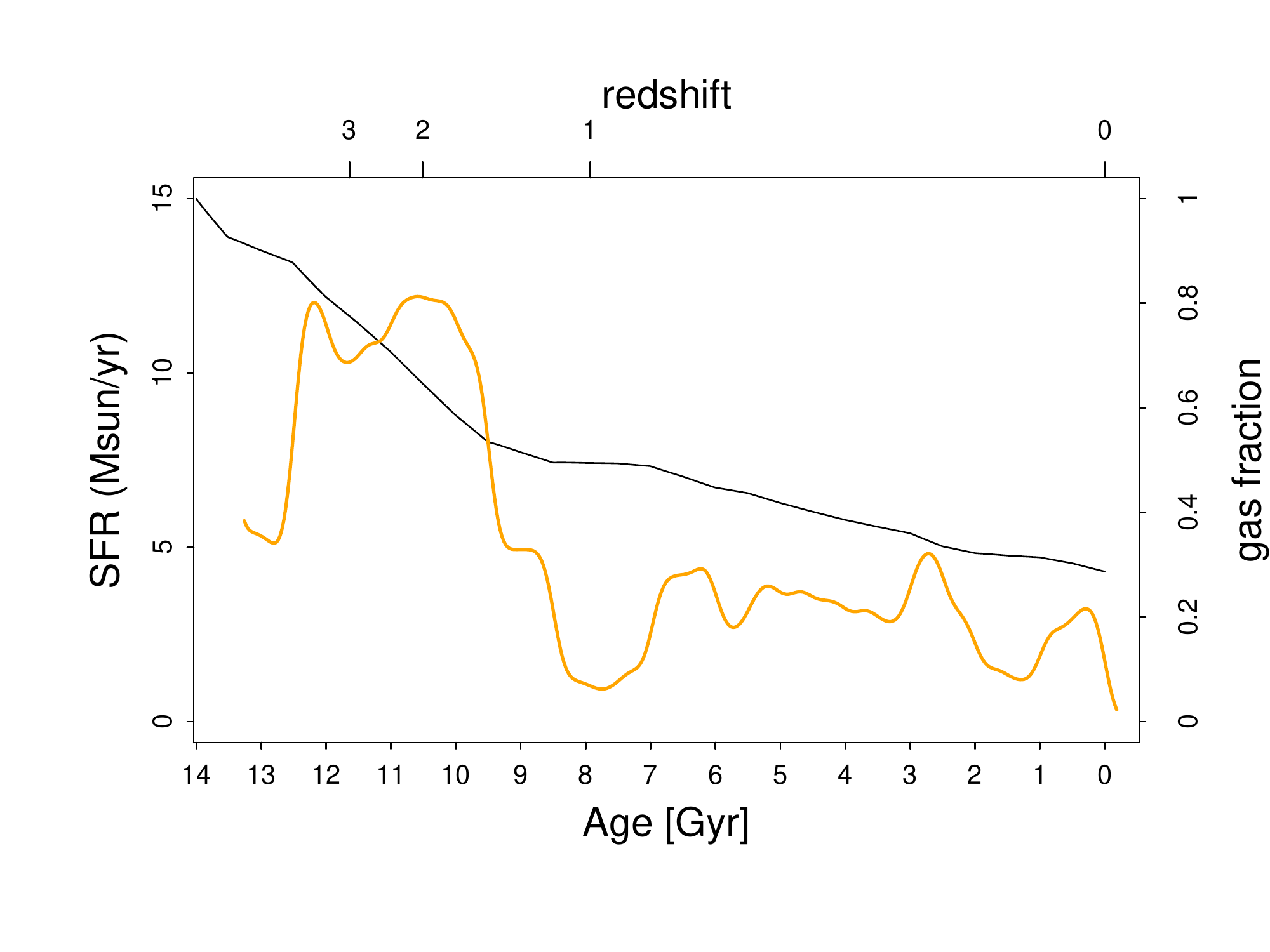}
\caption{Star formation history of the inner Milky Way (R$<$10kpc) derived from fitting the solar
vicinity age-[Si/Fe] abundance with a chemical evolution model in
\cite{snaith14, snaith15} (orange curve, left axis), together with the
gas fraction evolution in the model (thin black curve and right axis).
The SFR at $\sim$8 Gyr is negligible (consistent with no star formation),
while the gas fraction is still very high in the system, similar to the
molecular gas fractions estimated in disks at redshifts $\sim$1-3 
\citep{daddi10,aravena10,dannerbauer09,tacconi10,tacconi13}. 
The SFH is normalised such that the current stellar mass of
the Milky Way is 5$\times$10$^{10}$M$_{\odot}$.}
\label{fig:sfh}
\end{figure}

While several papers have suggested that the thick disk could be 
formed in a burst phase of star formation more than 10 Gyr ago
\citep[e.g][]{burkert92,chiappini99,fuhrmann04,reid05,bernkopf06}, we argued
in \cite{haywood13} that the thick-disk population extended from about 
13 to 9 Gyr ago. 
Indeed, the modelling of solar vicinity chemical 
abundances in \cite{snaith14,snaith15} has shown that 
the inner ($<$10kpc) disk\footnote{We use here the nomenclature 
introduced in \cite{haywood13}, Snaith et al. (2015), and Haywood et al. (2015).
The inner disk is made of stars that belong to both the thick and thin disks
and form the sequence that extends from metal-poor $\alpha$-rich abundances to the metal-rich $\alpha$-poor
sequence. It dominates the chemical patterns, as seen in the APOGEE data at R$<$7 kpc, see Fig. 3. 
This sequence is outlined by our model in both Fig. 2 for solar vicinity data and Fig. 3 for APOGEE
data. The outer disk is defined by the low-$\alpha$ sequence, visible in the solar vicinity chemical patterns
(see Haywood 2008), and now beautifully illustrated by the APOGEE survey with in situ stars.
} 
of our Galaxy shows two distinct phases
of star formation, see Fig. \ref{fig:sfh}.
The first corresponds to the formation of the thick disk between 13
and 9 Gyr, the second phase between 7 Gyr and the present epoch corresponds to
the formation of the thin disk.  At the end of the thick-disk phase,
between 10 and 9 Gyr, the star formation in the MW dropped
from a high-intensity starburst regime \cite[see also ][]{lehnert14}
to a more quiescent phase at z$<$1.  At this epoch, the decrease in
the star formation rate (SFR) was much faster than expected from the
exhaustion of gas as provided by a Schmidt-Kennicutt relation. This
implies that star formation ceased not due to a shortage of gas, but
because star formation was inhibited and particularly inefficient (i.e.
had very long gas depletion times).  Between $\sim$8.5 and 7.5 Gyr,
the star formation history (SFH) of the MW shows a prominent
dip, compatible with very low rates of, or no star formation during
this period.  Although galaxies are expected to experience fluctuations
in their SFR, this dip is made peculiar by marking the end of
the growth of the thick disk and the transition to another dynamical
and morphological stellar population.  In the present study, we argue
that this cessation of the SFR corresponds to star formation quenching
that is observed in galaxies at high redshifts.  Using studies of the
demographics of disks with mass lower than $M^*$, the characteristic
stellar mass of the galaxy stellar mass function, in the local volume,
we argue also that quenching seen in Milky Way-like disks galaxies and
later types at high redshifts is not bulge building \citep[e.g.][]{abramson14}, but represents the end of thick-disk building.

\cite{snaith15} found the dip in the SFH by fitting a model to the
age-[$\alpha$/Fe] relation of solar vicinity stars. Because the dip
occurs at the end of the thick-disk phase, an epoch when stars tend
to form on more circular orbits, it was not clear whether this feature was only
localized around the solar circle or if it was more pervasive, happening
throughout the galactic disk. Orbits becoming more circular means that
stars from the inner disk may not be fairly represented at the solar
circle, and thus any features in the SFH are only constrained for the
evolution of the MW at around the radius at which the Sun lies from the
Galactic centre.  
It is important at this stage to emphasize that the SFH was derived from 
fitting the age-[Si/Fe] relation obtained from an inhomogeneous and incomplete
sample. However, since we did not rely on stellar densities, this has
no or limited impact because as far as the abundance patterns observed in 
Adibekyan et al. are correct and representative, the age-[Si/Fe] will be a
faithful record of the past star formation rate. We also note that \cite{nissen15} 
confirmed the existence of a tight correlation between $\alpha$
elements and age in the thin-disk regime.
Indeed, that the abundance patterns of the Adibekyan et al. sample  are correct and 
representative can be checked by comparing with similar samples \citep[e.g.][]{bensby14} 
or, more extensively, on the APOGEE data \citep{majewski15}, Fig. 3.
Therefore, the SFH derived in \cite{snaith15} is valid for the whole inner disk,
as far as the stars that enter the age-[Si/Fe] relation are representative of the 
whole disk. Derivation of the orbital properties of these stars in \cite{haywood15}, Fig. 4, 
which shows the pericenters of stars as a function of their $\alpha$ abundance and age,  illustrates
that this must be the case for stars older than about 8-9 Gyr (thick-disk stars), which reach
the innermost regions of the disk, but it must be checked for younger stars.

The aim of the present paper is therefore to confirm the validity over the whole disk of a SFH
measured locally, and in particular that of the dip seen between 9 and 7 Gyr,
by comparing our model with star counts several kpc
from the Sun. We demonstrate that this dip is visible throughout the inner disk
(R$<$10kpc) -- it is a general feature in the global evolution of the
SFH of the MW.

The outline of this paper is as follows: in the next section, we briefly
describe our model and discuss the evidence that the star formation in our
Galaxy ceased about 8 Gyr ago from spectroscopic and age data of a sample
of stars in the solar vicinity.  In Sect. 3 we present our selection
of the APOGEE data and the comparisons with our model.  We discuss these
results in Sect.~\ref{sec:discussion} and in particular its significance
for the evolution of Milky Way-type galaxies generally. We provide what
we conclude from this study in Sect. 5.

\section{Model and evidence from the solar vicinity}

The interpretation of the solar vicinity chemical patterns is paved with a rich 
literature \citep[][]{matteucci89,gilmore91,timmes95,chiappini97,haywood01,schonrich09,kobayashi11}.
\cite{haywood13} emphasized that the low-$\alpha$ and $\alpha$-rich
sequences in the [$\alpha$/Fe]-[Fe/H] plane (see Fig.~2) are possibly the
result of two different chemical evolutionary histories, those of the
inner and outer disks. We can sample these two sequences at the solar
vicinity because the Sun is at the interface between them. These two
chemical evolutions, having widely different metallicities at similar
ages, cannot be represented by a single model \citep{haywood13}.
We emphasized that while the inner-disk sequence is typical of
a closed-box evolution, the lower-$\alpha$ sequence is representative of chemical
evolution with dilution. These models were described and compared
to chemical trends in the solar vicinity in \cite{snaith15}.
\cite{halle15} proposed that the outer Lindblad resonance
(OLR) of the bar, located in the solar vicinity \citep{dehnen00}, is responsible for keeping these two parts of the disks relatively
well separated.  APOGEE data have confirmed these results by showing
that the inner disk essentially lacks low-$\alpha$ stars, while
the outer disk is dominated by the low-$\alpha$ sequence \citep[e.g.][]{hayden15}.  The SFH we derived
in \cite{snaith15} is the general SFH of the inner disk (R$<$10kpc),
and the quenching event described here applies to the inner disk only,
not the outer disk. As shown in \cite{snaith15}, the outer disk (beyond
the OLR) is characterized by different chemical patterns that translate
into a different, flatter, SFH. Hereafter we analyse the inner-disk sequence
only, which extends from the $\alpha$-rich metal-poor to the $\alpha$-poor
metal-rich stars. Although this sequence is often called the thick-disk sequence, we prefer to call it the {\it \textup{inner-disk}} sequence
because we think it describes the evolution of the whole inner disk, comprising
both the thick and thin disks 
(essentially pure inner disk at R$<$7kpc, and a mix of it with the outer disk, 
between 7 and 10 kpc, the OLR region).

\subsection{Chemical evolution model and derived SFH}

The level of [$\alpha$/Fe] abundances at a given time is determined by the
ratio of the present to past SFR \citep{gilmore91}. The higher the past SFR (relative to the present SFR), the 
lower the [$\alpha$/Fe] abundance at more recent times, implying a steep decrease
in [$\alpha$/Fe] abundance, and a steeper slope in the age-[$\alpha$/Fe] abundance relation. 
The slope of this relation  reflects past SF intensity.
This is illustrated by the SFHs of Fig. 4a and the corresponding age-[$\alpha$/Fe] relations
of Fig. 5.
The information on the SFH of the disk is therefore fossilized in the run of $\alpha$-element
abundances with stellar age in the solar vicinity \citep{haywood13}
and has been used in \cite{snaith14,snaith15} to derive a general
history of the star formation in the MW by fitting a chemical
track to the age-[Si/Fe] relation.  To fit the age-[$\alpha$/Fe]
relation, a chemical evolution model must be assumed. We adopted a
closed-box model for the reasons given in \cite{haywood14a,haywood14b}
and \cite{snaith15}.
We emphasized
that the closed-box is taken here to mean a model where most accretion
has occurred early (before substantial star formation occurred) in the
inner disk, motivated by expectations that infall in the central regions
of galaxies must be rapid \citep{haywood15}.  \cite{snaith14,snaith15}
did not assume any dependency of the SFR on the gas density,
as is typically done in chemical evolution models, but derived the SFH
that allows the model to best fit the age-[Si/Fe] relation (see also
Fig.~\ref{fig:sfh}). Detailed accounts of the procedure used to fit the
stellar age-[Si/Fe] relation, its robustness and uncertainties of the
method and result were given in these two papers. 
Detailed comparisons between our model and the observed MDF from 
APOGEE are the focus of a forthcoming article.

Figure~\ref{fig:sfh} also shows the evolution of the gas fraction as a
function of time by our model.  It shows that at the end of the thick
disk phase (9.5 Gyr, z=1.5), the amount of gas in the system is still
very large ($\sim$50\%). Similar fractions have been reported on massive
galaxies at these redshifts, see for instance \cite{daddi10} or
\cite{bethermin15}, but it remains to be confirmed on MW-type
galaxies\footnote{This is also compatible with the gas fractions that are
implied by the derived SFHs for MW-type galaxies. For instance, \cite{papovich15}
inferred that while the gas fraction in M31-type galaxies has decreased to lower than 20\% at z$\sim$1.5, it is still high for MW-sized
galaxies at $\sim$40\%, consistent with the fraction estimated here. See also
\cite{popping15}.}.

The model is able to follow the evolution of silicon, magnesium, and
oxygen. Because of problems with magnesium theoretical yields, however, it
is difficult to compute a mean $\alpha$ abundance from these elements to
compare to the APOGEE data. Silicon theoretical yields are well behaved
and have been used to constrain our model. However, the silicon abundances in the APOGEE
data are significantly different from the silicon abundance determined for local samples of stars. To avoid being trapped in the details of the behaviour of one
particular element, we decided to compare mean $\alpha$ abundances. To do so,
we converted the silicon model abundance to [$\alpha$/Fe] using a relation
obtained from the observed [Si/Fe] and [$\alpha$/Fe] of the \cite{adibekyan12} sample. The model was then
compared to the data (Fig. \ref{fig:alphafehsolar}). This is justified by
the surprisingly close similarity between the behaviour of the $\alpha$
solar vicinity abundance scales and from APOGEE, as corroborated by the
distributions of Fig. \ref{fig:alphafehapogee}, although they are derived
from different data and elements. The comparison of the [$\alpha$/Fe]
distributions between the two samples shows that the systematic shift
between the two is at most 0.02 – 0.03 dex and is thus insignificant for
our analysis.

\subsection{What do we wish to measure?}

\begin{figure}
\includegraphics[trim=40 120 0 80,clip,width=9.cm]{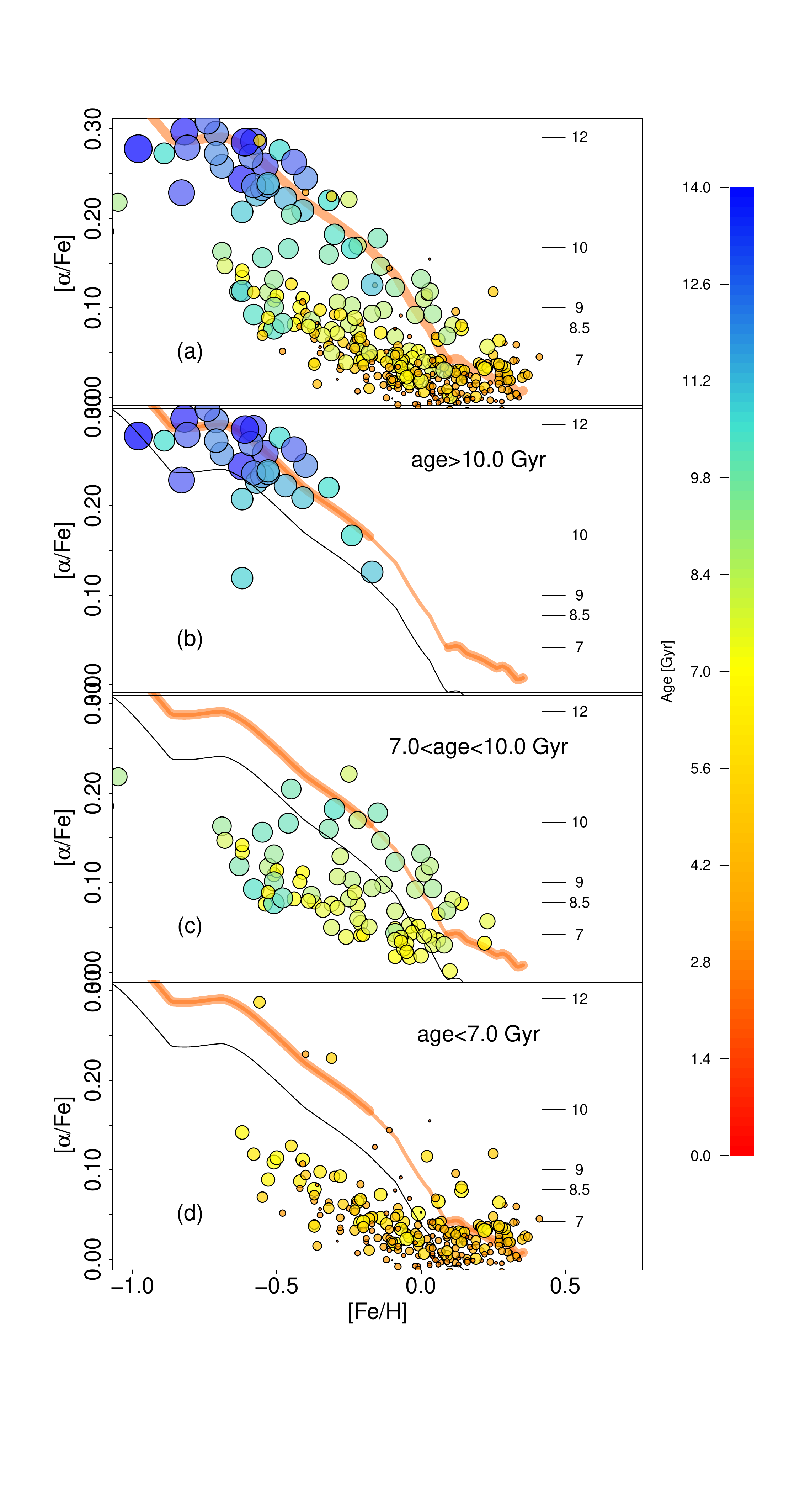}
\caption{ [$\alpha$/Fe]-[Fe/H] distribution at the solar vicinity
for the sample of \cite{adibekyan12} with ages from \cite{haywood13}.  
The colours (indicated in the legend) and size of the 
symbols indicate the age of the stars (larger
circles imply older ages).
The orange line represents the model of  \cite{snaith15},
which has the SFH of Fig. 1. The thinner segment of the line 
corresponds to the lull in the SFH of Fig. 1, between 10 and 7 Gyr ago.
The thin black curve is the same model shifted down by -0.05 dex and is used
in Sect. 3 to select inner-disk stars from the APOGEE survey.
The stars below and to the left of this line are either accreted stars
according to Nissen \& Schuster (2010; for those objects which have
[Fe/H]$<$-0.7 dex), or metal-poor thin-disk stars.
A vertical scale within each plot indicates ages along the model from 7 to 12 Gyr.}
\label{fig:alphafehsolar}
\end{figure}

Figure~1 shows that the SFR of the MW  thick and thin disks at R$<$ 10 kpc has dropped suddenly 10 Gyr ago to
reach a minimum $\sim$ 8 Gyr ago.  As mentioned previously, this event has
been found in \cite{snaith14} by fitting the age-[Si/Fe] relation.
We now wish to search for its signature in the star counts. Because of
the tight correlation that exists between age and [$\alpha$/Fe] \citep{haywood13,haywood15}, we expect any variation in the SFH to be imprinted
on the number density of stars as a function of [$\alpha$/Fe] and to a
minimum in the SFH should correspond a minimum, a dip, or a bimodality in
the stellar counts as a function of [$\alpha$/Fe].  Now, we must specify
where we wish to measure the variation of densities in [$\alpha$/Fe].
When slicing the [Fe/H]-[$\alpha$/Fe] plane at constant [Fe/H] (i.e.,
vertically), a bimodality arises because we successively 
cross the relatively $\alpha$-rich and then $\alpha$-poor sequences. 
The chemical evolution from the thick to the thin disk has been described as 
parallel tracks connecting these two sequences (e.g. Sch\"onrich \ Binney 2009,
their Figs. 3 \& 4).  In our view, however, for reasons given in the
previous section, the chemical evolution does not connect the two sequences.
The bimodality that arises from the dip that
separates the $\alpha$-rich sequence from the metal-poor thin-disk sequence
when slicing the  [Fe/H]-[$\alpha$/Fe] plane vertically has no 
evolutionary meaning (see \cite{nidever14} or \cite{kordopatis15} for a characterization 
of the distance that separates the two sequences).

The evolution of the inner disk occurs along a single well-defined
sequence in the  [Fe/H]-[$\alpha$/Fe] plane, which reaches from metal-poor
$\alpha$-rich ([$\alpha$/Fe]$\sim$0.3 dex) stars to $\alpha$-poor metal-rich
objects ([$\alpha$/Fe]$\sim$0. dex, [Fe/H]$>$0.0 dex).  This sequence
is well defined observationally in the data of \cite{adibekyan12},
see our Fig. \ref{fig:alphafehsolar}, and well reproduced by our model (orange curve).  
The bimodality in which we are
interested in here is measured {\it \textup{along}} this sequence and not
at constant [Fe/H].

Figure~\ref{fig:alphafehsolar} shows the [Fe/H]-[$\alpha$/Fe] distribution
for a sample of F and G dwarfs from \cite{adibekyan12} for which
ages could be determined \citep{haywood13}.  The same stars have
been used to derive the SFH in Snaith et al. (2014, 2015, and Fig. 1)
by fitting the age-[Si/Fe] relation.  The [Fe/H]-[$\alpha$/Fe] chemical track
corresponding to this SFH has been overplotted on the data. We note
that it is not a fit to the data, but results directly from the fit made
from the age-[Si/Fe] relation.  We converted  [Si/Fe] to [$\alpha$/Fe]
using a regression fit between these two quantities for stars in the
Adibekyan sample.  This model corresponds to the inner-disk sequence and
is shown as a continuous line from the old $\alpha$-rich stars to
the metal-rich $\alpha$ poor objects.  We emphasize that this model describes
the evolution of the inner-disk sequence alone.
It does not represent the lower-$\alpha$ sequence with [Fe/H]$<$0.0 dex.
The black curve in Fig.~\ref{fig:alphafehsolar} is the inner-disk sequence shifted by -0.05 dex.
The inner-disk stars were chosen such that they are above this limit.
In the following sections, we use this criterion to generally
select the inner-disk stars.

The age scales in Fig. 2 -- both the observational scale as illustrated
by the symbols and the model scale --  indicate that the inner-disk
sequence is a temporal sequence.  Figure~\ref{fig:alphafehsolar}, panels b,
c, and d show the [$\alpha$/Fe]-[Fe/H] abundance distribution dividing the
sample in three different age intervals.  The model track is separated into
three segments, with the one between 10 and 7 Gyr corresponding to the
lull in the SFH of Fig. 1; it is represented by a thinner line.  Hence,
we expect that a pause between 7 and 10 Gyr in the SFH corresponds
to a significantly lower density of stars between about 0.05 and 0.17 dex.

There is indeed a hint of a minimum density in this interval when we plot
the histogram of [$\alpha$/Fe] for stars in the \cite{adibekyan12}
sample (see Fig. 3a).  Given the nature of the sample used here, however,
which is both not complete over its volume and has limited numbers
of stars, any dip in the histogram is difficult
to discern.  Because of these limitations, we now turn to the APOGEE
survey with its higher completion and significantly better statistics
\citep{holtzman15}.
Implicit in the present work is the assumption that the age-[$\alpha$/Fe] 
relation deduced from stars in the solar vicinity is valid over the inner (thin and thick) disk.
This assumption is based on our previous study \citep{haywood15},
where we have shown that the tight correlation observed in the solar vicinity 
between age and $\alpha$ abundances strongly favours a uniform SFH over the entire radial distribution 
of the inner disk. 
Haywood et al. (2015, see their Fig. 4) showed that the pericenters of the thick-disk stars in the sample
reach the innermost Galaxy (pericenters $<$ 2 kpc). Hence, the sample, although local, 
probes the entire inner disk.

As mentioned at the beginning of this section, the speed at which [$\alpha$/Fe] abundance
varies with time is  a function of the star formation intensity. Because of the wide orbital range 
of the stars sampled in the solar vicinity, the observed tight correlation between age
and  [$\alpha$/Fe] abundance supports a homogeneous SFH in the inner disk.

\section{Archeology of a quenching event in the Milky Way star formation
history: stellar densities}\label{sec:quench}

\subsection{Selected data}\label{sec:data}

Starting from a sample of 83168 stars with distances, our selection of APOGEE objects is similar
to the one made by the APOGEE collaboration \citep{hayden15}: we
chose objects with T$_{\rm eff}$$<$5500K and 1.0$<$log~g$<$3.8. Since
the determination of [$\alpha$/M] may be less reliable \citep{holtzman15} below 4000K, we also removed objects below this limit, amounting
to 35965 stars between 3 and 9 kpc from the Galactic centre.

Distances derived for the whole APOGEE survey were taken from \cite{schultheis14}. They have been
derived using the Padova set of isochrones. For each star, the closest
point on the isochrones was searched for in the [M/H], log~g, T$_{\rm
eff}$ space. Stars for which a point on the grid closer than 500K in
temperature or 0.5 dex in log~g could not be found were discarded. In addition,
3\% of the stars do not fall within the grid of isochrones and were
also discarded. Absolute magnitudes in the 2MASS photometric bands
were estimated from this procedure and were used to estimate both the
extinction and the distance of each star. The median error in distance
estimates, taking into account the uncertainties on the atmospheric
parameters from APOGEE, is of about 30-40\% \citep{schultheis14}\footnote{The distances were also compared by one of the authors (M. Schultheis) to Hayden et al. (2014) and Santiago et al. (2015),
and no systematic difference was found.}.
Relying on red clump stars only would have provided 
more accurate distances. However, the  APOGEE-RC  catalogue \citep[see][]{bovy14}, containing 10341 objects, with the 
majority at Galactocentric distances larger than 8~kpc, does not sample the inner disk particularly well.

\begin{figure}
\includegraphics[trim=40 10 0 10,clip,width=9.cm]{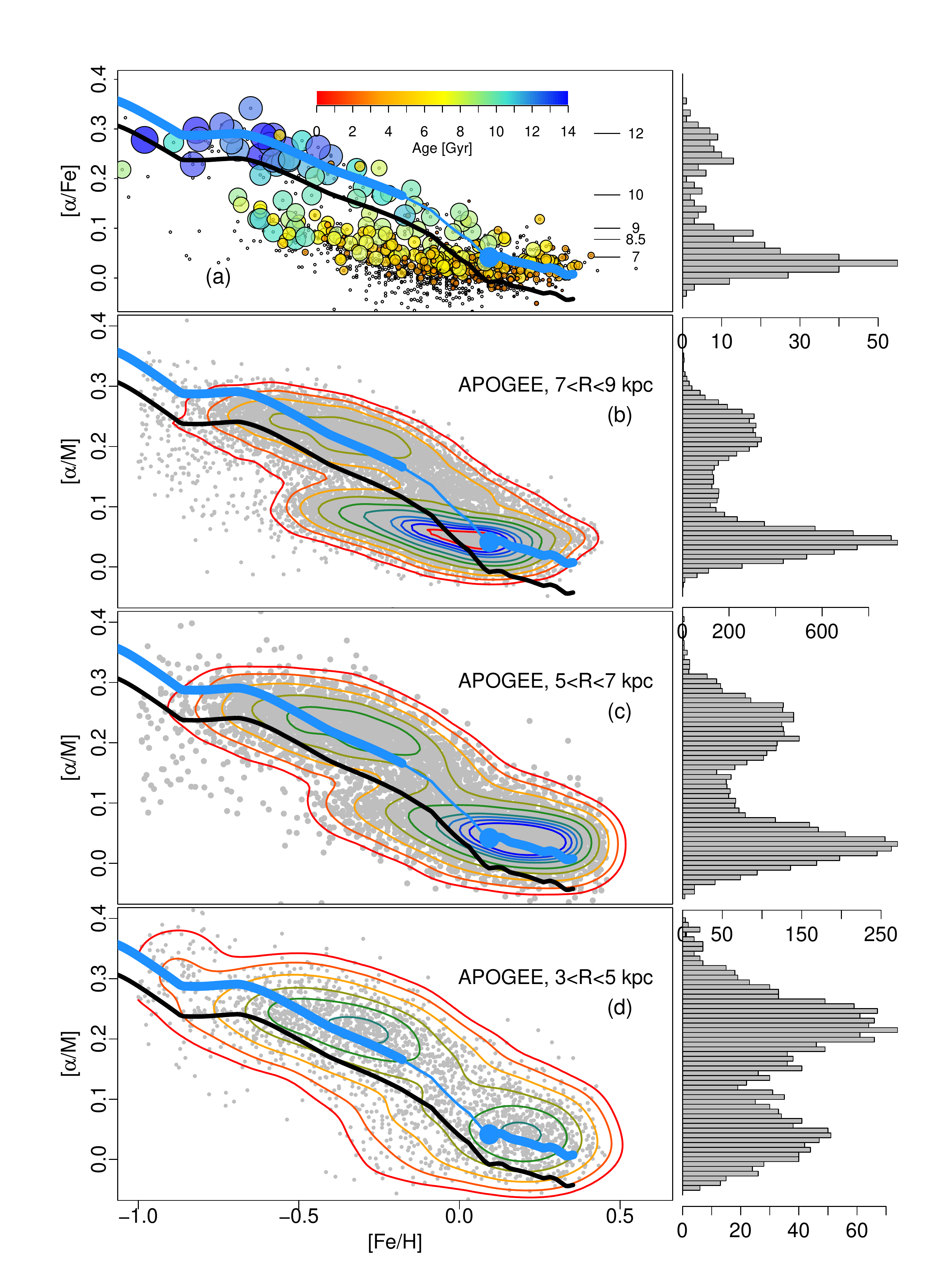}
\caption{(a) Abundance distributions for stars in the solar
        vicinity from very high resolution spectroscopic data and (b), (c), and (d) data from
APOGEE. The inner-disk sequence (above the black line) is composed of thick-disk and metal-rich thin-disk stars and dominates the inner-disk
stellar populations, as illustrated in (c) and (d). The blue thick
curve represents the track of our chemical evolution model, with the dot indicating
the beginning of the thin-disk era at 7 Gyr. The thinner blue segment indicates the quenching phase
from 10 to 7 Gyr.
The black line is defined as the model$-$0.05 dex in [$\alpha$/Fe]. 
We emphasize that the model (blue line)
represents the evolution of the inner-disk sequence alone (stars above the black line). 
The outer-disk sequence (below the black line) represents a different evolution 
and requires a different model (see Snaith et al. (2015)).
The histograms
count the number of stars above the black line. In (a), grey points
are all the stars in the solar vicinity sample, while coloured circles
are a subsample for stars with age. The size and colour of the symbols
code the age of the stars. The tick marks indicate the age of the model
along the track. The histogram includes all stars above the black line
(whether they have an age determination or not).}
\label{fig:alphafehapogee}
\end{figure}

We now apply the same selection as for the sample of solar vicinity stars,
see Fig. \ref{fig:alphafehapogee}. The blue curve represents the model
of the inner galactic disk described in the previous section. The black
curve is the same model lowered by 0.05 dex, and we select stars with
a good probability of belonging to the inner-disk sequence as those stars
above this limit. This limit is somewhat arbitrary, but it was chosen to
select stars on the $\alpha$-rich, inner-disk sequence and exclude those on
the $\alpha$-poor (outer) thin-disk sequence. 
It must be noted that solar metallicity stars that belong to the low-$\alpha$
sequence are essentially discarded by the selection, and the Sun itself would not be
selected.  The model track in Fig. 3 shows that the inner disk had already
reached solar metallicity 9 Gyr ago, or roughly 4 Gyr before the solar vicinity.
This means that solar vicinity stars, or OLR stars, have followed a slightly different 
chemical evolution history than inner-disk stars.
Hence, although the quenching episode we discuss in the present study  occurred 
when the inner disk had reached solar metallicity, solar vicinity stars of 
solar metallicity were not affected because they formed $\sim$ 5 Gyr later. 

Therefore, stars selected in the solar
vicinity (Fig. \ref{fig:alphafehapogee}b) are likely to be contaminated
by objects born at the interface between the inner- and outer-disk stars that are not truly members of the inner disk, such as the Sun.  Histograms on the
right show the number of selected stars in bins of [$\alpha$/Fe] once
the outer disk is removed. They all show a bimodal distribution with
a minimum between 0.17 and 0.07 dex. We emphasize that the minimum does
not separate the thick disk from the thin disk, which were defined as those
stars that make the low-$\alpha$ sequence.  The minimum is a variation of
stellar densities along the inner-disk sequence and separates the thick
disk from the thin disk, defined as stellar populations in the sense given
in \cite{haywood13}.  In this sense, the two populations are defined
by distinct segments in the age-[$\alpha$/Fe] relation, which have been
shown in \cite{snaith15} to correspond to two distinct phases of
star formation. With this definition, the thick and thin disks follow
the blue track in Fig.  \ref{fig:alphafehapogee}, with the thin disk
represented by the portion to the right of the dot on the blue track.
At all distances from the Galactic centre, the $\alpha$ distribution is
clearly bimodal and shows a clear dip between 0.07 and 0.17 dex.

It has been shown by \cite{hayden15} that the sampling in directions,
magnitudes, and colours of APOGEE does not introduce any significant bias
in the metallicity distribution function of the survey.  Giants, however,
probably bias the underlying age distribution against the oldest
objects or the most $\alpha$-rich stars, see \cite{bovy14}.  Hence,
we keep in mind that the peak at  [$\alpha$/Fe]$\sim$+0.23 dex possibly
underestimates the relative number of old stars compared to younger ones.

\subsection{Model stellar densities: why is the [$\alpha$/Fe] distribution
bimodal?}

\begin{figure}
\includegraphics[trim=40 360 0 80,clip,width=9.cm]{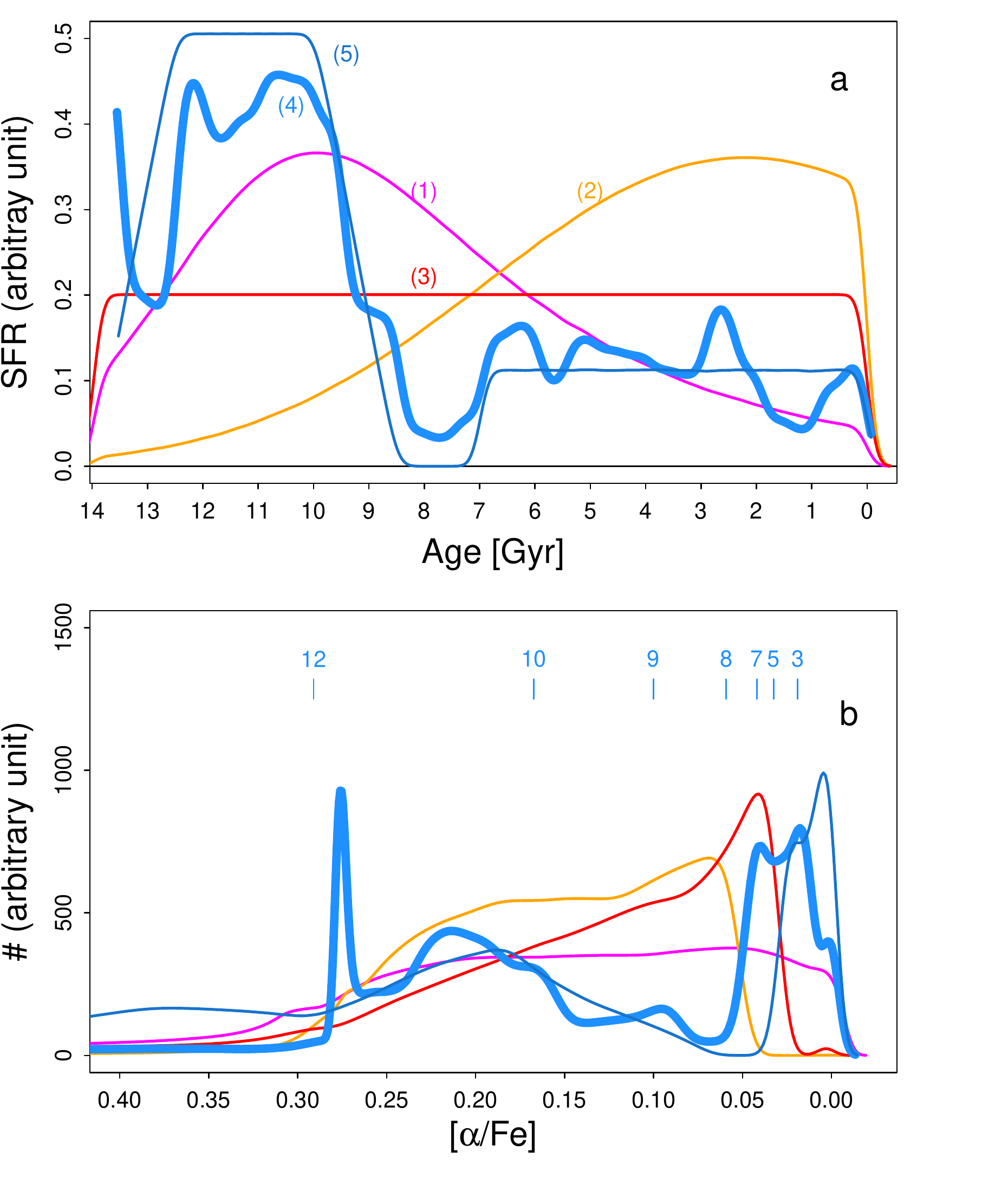}
\caption{
{\bf a)} Broad range of SFHs designed to cover
decreasing, increasing, and constant star-formation rates with time
(labelled 1, 2, 3), and the SFH from \cite{snaith14, snaith15},
with a dip between 8.5 and 7 Gyr (thick blue curve labelled 4). SFH
5 is similar to 4 with noise removed (thin dark blue curve). {\bf b)} The
corresponding stellar densities as a function of [$\alpha$/Fe]. SFHs
like 1, 2, and 3 do not generate a dip in the [$\alpha$/Fe]
distribution. Only an SFH like 4 (or 5) with a drop and cessation in
star formation in the specific time interval between $\sim$10 to 7 Gyr
can generate a distribution with a significant dip (blue curve). The
short blue vertical lines indicate the time variation in Gyr for model 4
(thick blue curve). The peak at [$\alpha$/Fe]$\sim$0.28 dex is due to
the flattening of the variation of [$\alpha$/Fe] visible in Fig. 1,
which is not at all significant compared to the data used to generate
the SFH \citep{snaith14, snaith15}.}
\label{fig:alphamod}
\end{figure}

\begin{figure}
\includegraphics[trim=40 10 0 20,clip,width=9.cm]{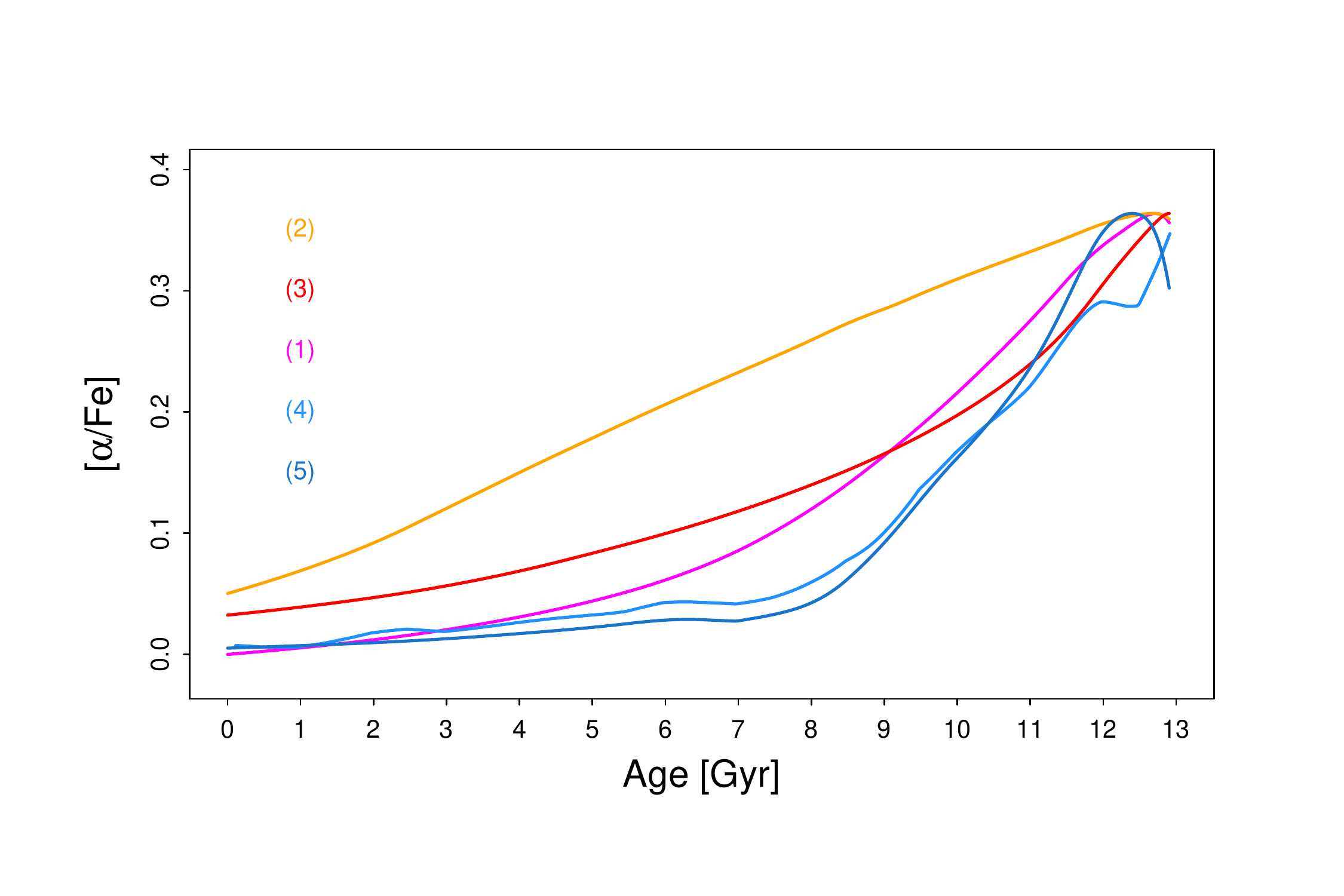}
\caption{Age-[$\alpha$/Fe] relations generated by the various SFHs of Fig. 4a. Numbers correspond
to the SFHs of Fig. 4a. The slope
of the age-[$\alpha$/Fe] relation at a given age is proportional to the intensity of the SFR at this age.}
\label{fig:alphaage}
\end{figure}

The bimodality of the  [$\alpha$/Fe] distribution has been discussed by others, 
see for instance \cite{fuhrmann04},  \cite{schonrich09}, and \cite{bovy12a}.
We now show that a bimodal density distribution along the inner-disk sequence
of [$\alpha$/Fe] requires a very specific form of the SFH. To
illustrate this, we first calculate the expected density distributions in
[$\alpha$/Fe] for a range of SFHs along with their corresponding density
distributions (Fig.~\ref{fig:alphamod}).  The age scale shows the temporal
evolution of the [$\alpha$/Fe] abundance ratio for the SFH 4.  Two results
are worth noting.  First, different SFHs lead to different minimum
[$\alpha$/Fe] abundance ratios at the end of the evolution.
SFH 2 for example, with a slowly rising
SFR, produces only a limited number of SNIa at early times and therefore
a limited amount of iron at later times, hence the [$\alpha$/Fe] stays
high -- reaching values slightly below 0.05 dex.  This is compared to
SFHs 1 and 4, which have an SFR sufficiently high at early times such
that the iron subsequently generated decreases the [$\alpha$/Fe] ratio
just below 0.  SFH 5 reaches even lower [$\alpha$/Fe] values because of a
slightly higher SF intensity between ages of 12 and 10 Gyr. This is also
illustrated in Fig. \ref{fig:alphaage}, which shows the age-[$\alpha$/Fe]
relation for the various SFHs in Fig. \ref{fig:alphamod}.

Second (and most important) for SFHs 1, 2, and 3, the corresponding stellar
densities reach a maximum, or near maximum density, in the interval
where the data show a minimum density, between 0.05 and 0.17 dex in
[$\alpha$/Fe].  The SFH from Snaith et al. (2014, 2015) is plotted in
Fig. \ref{fig:alphamod}a and the corresponding stellar densities as
a function of [$\alpha$/Fe] in Fig. \ref{fig:alphamod}b, SFH 5
is equivalent to SFH 4 (except for a higher intensity between 12 and
10 Gyr), with the noise smooth from the SFH (4).

Star formation histories 4 and 5 show that a sudden drop and cessation of the star formation
activity between 7 and 8.5 Gyr are able to generate the dip in the stellar
densities observed in the data. Figure \ref{fig:alphamod} demonstrates
that the observed shape of the [$\alpha$/Fe] distribution requires a
particular SFH. Because of the tight correlation between [$\alpha$/Fe] abundance ratio
and stellar age, the drop in the SF activity must occur at a precise
and well-constrained epoch.  

The bimodality in the $\alpha$ distribution arises for two reasons.
The first reason is the slow evolution of [$\alpha$/Fe] with age in the
thin disk, which is reflected in the shallow slope of the age-[$\alpha$/Fe] relation at
$<$7 Gyr (see Fig. \ref{fig:alphaage}).  For ages younger than this,
the variation of [$\alpha$/Fe] with age becomes slow and stars accumulate around [$\alpha$/Fe]$\sim$+0.025 dex, as is the case for all SFHs of
Fig. \ref{fig:alphamod} except for SFH 1. 
The steeper slope of the age-[$\alpha$/Fe]
relation in the thick-disk sequence has the effect of decreasing the number
of stars per unit [$\alpha$/Fe]. Together with the lull in the SFH at 8
Gyr, this generates the dip in the [$\alpha$/Fe] distribution
between 0.07 and 0.14 dex.  The stellar density rises again only towards higher [$\alpha$/Fe]
as a result of the higher SFR during
the thick disk phase, generating the bimodality that is seen in
Figs. \ref{fig:alphafehapogee} and \ref{fig:alphasfr}.
This means that the bimodality (and the dip in between) is the result of both the SFH and the nature of the age-[$\alpha$/Fe]
relation.

\subsection{Model-data comparisons}

\begin{figure*}
\includegraphics[trim=10 10 0 10,clip,width=9.cm]{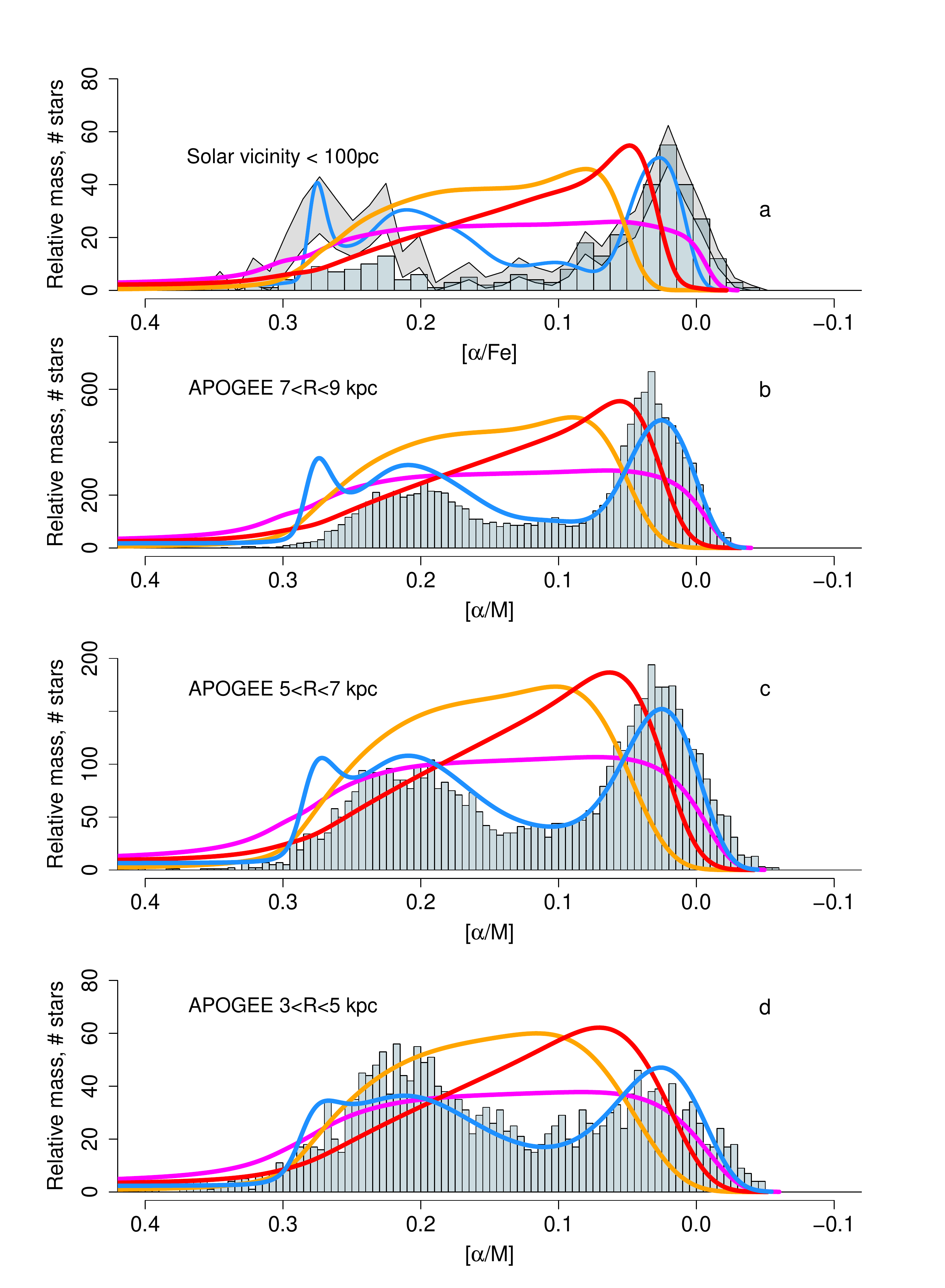}
\includegraphics[trim=10 10 0 10,clip,width=9.cm]{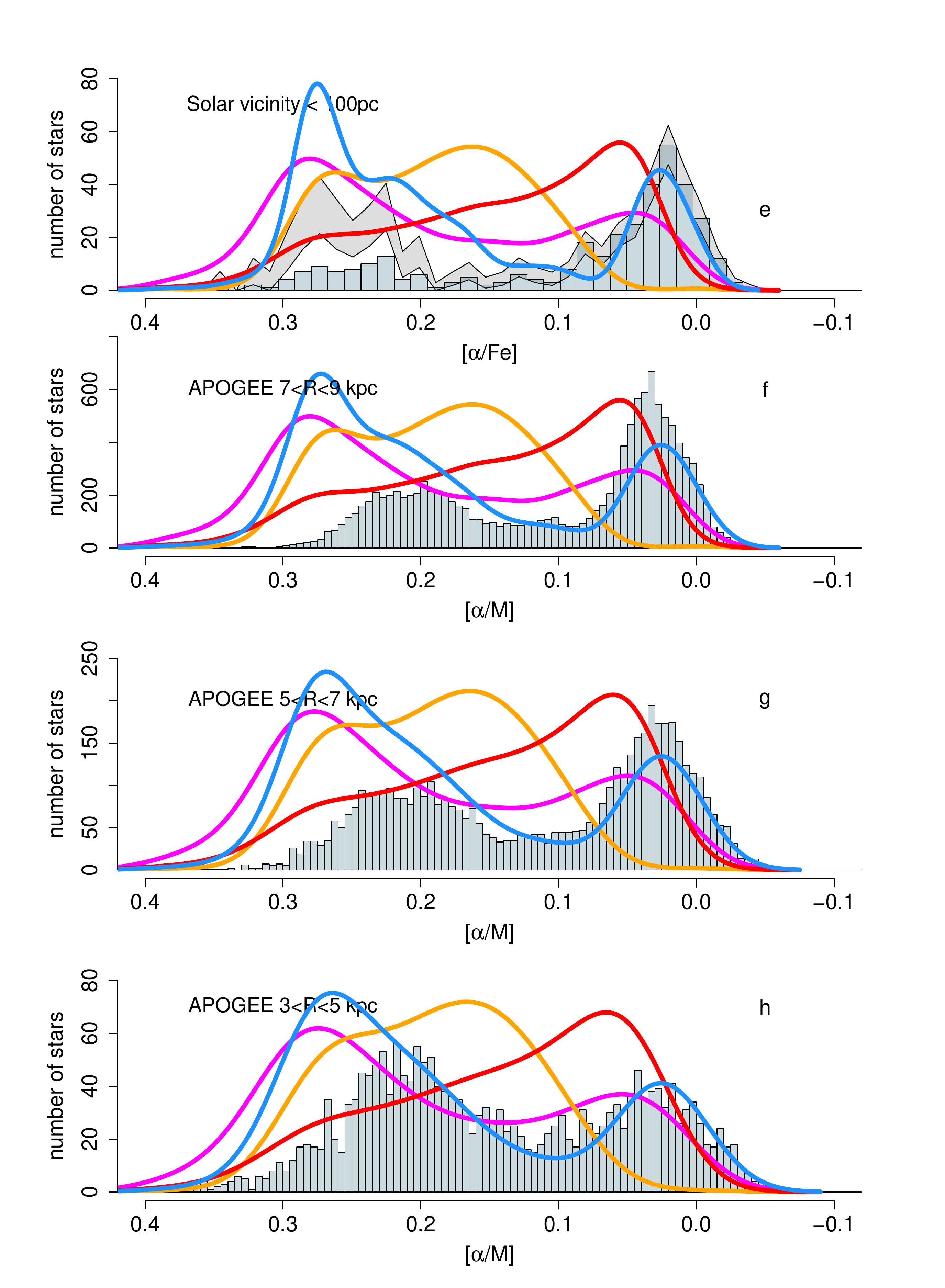}
\caption{(a) Comparisons between observed and modeled mass (abcd) and star number (efgh) distributions as a function of
[$\alpha$/Fe]. Plots (a) and (e) show our calibration sample (solar vicinity) and panels (bcd) and (fgh) 
the APOGEE survey. The solar vicinity observed distribution in panels (a) and (e) has been corrected to
account for the lower sampling of old populations in the Galactic
plane. From top to bottom, models have been convolved with increasing
dispersion (0.01, 0.015, 0.020, and 0.025 dex) to account for the
difference in data quality. For comparison with APOGEE data, models have been
shifted by $+$0.025 dex to allow for an offset in the $\alpha$-element abundance
scale between our calibration sample and APOGEE.}
\label{fig:alphasfr}
\end{figure*}

Figure \ref{fig:alphasfr} shows both the stellar mass (a to d) and the stellar counts (e to h) as a function of
[$\alpha$/Fe] or [$\alpha$/M] 
of all SFHs compared to the histograms of the stars belonging to the inner-disk sequence
of the four datasets used to make Fig. 3.  
The stellar counts were evaluated by assuming a Kroupa (2001) IMF (the same as assumed
in the chemical evolution model). At each age, we assumed that the mass is distributed according to the IMF along
an isochrone at the metallicity given by the age-metallicity of the model. For each isochrone, we selected stars that,
having log~g $<$3, are distributed along the giant branch. 
Admittedly, this is a rough procedure, but it aims to show that by taking  the effect of 
the IMF and the expected distribution of stars along the giant branch into account, our results are essentially the 
same. 
Stellar counts  (Fig. \ref{fig:alphasfr} e,f,g, and h)  show distributions that are generally more heavily weighted
in favour of $\alpha$-rich, older stars: younger giants, which are more massive, are 
less numerous, as expected from the IMF. This is particularly true for SFH 2.
The consequence is that, at [$\alpha$/M]$>$0.2 dex, even our best model lies significantly above the data.
However, this may well be an effect of using giants as tracers, which, as mentioned in  Sect. \ref{sec:data}, may have biased
the observed counts against $\alpha$-rich stars.
We note  that in spite of the crude conversion into counts and this possible bias, 
the bimodality of the distribution is preserved for SFH 4.

The data are confined to stars
within the solar vicinity (distances within 100pc) in Fig. 6a
and e. This systematically underrepresents old populations,
which are more widely dispersed throughout than younger populations
of stars. We applied a
correction to convert local stellar volume densities (grey histogram) into
relative surface densities (grey area, which reflects corrected densities
$\pm$ Poisson uncertainties) using the result that mono-abundance
subcomponents are isothermal \citep{bovy12c}.  The disk can therefore
be modeled as a series of exponential subcomponents of different ages,
or equivalently, [$\alpha$/Fe] abundances.  Assuming exponential density
laws in z, the corrections to relative surface densities only reflect
the scale height of each subcomponent.  Scale-height information for
subcomponents in the [Fe/H]-[$\alpha$/Fe] plane were derived by \cite{bovy12b} from SEGUE.  We used this information in the following way:
$\alpha$-rich stars (on the SEGUE scale) with [Fe/H]$<$$-$0.7 dex have
a scale height of h$_{\rm z}$=856pc, while those with [Fe/H]$>$$-$0.7
dex but [Fe/H]$<$$-$0.3 dex have h$_{\rm z}$=583pc. A metallicity of
[Fe/H]$\sim$$-$0.7 dex along the inner-disk sequence corresponds to
[$\alpha$/Fe]$\sim$0.25 dex on our scale, while a metallicity of
[Fe/H]$\sim$$-$0.3 dex corresponds to [$\alpha$/Fe]$\sim$0.2 dex.
Higher metallicities are divided into two groups: the
$\alpha$-rich and $\alpha$-poor ([$\alpha$/Fe]$<$0.10--0.15 dex),
with scale heights of 348 pc and 239 pc.  Normalized to this last
scale height, the corrections are 1.00, 1.45, 2.44, and 3.58 applied in
the intervals of [$\alpha$/Fe]$<$0.10, 0.15$<$[$\alpha$/Fe]$<$0.20,
0.20$<$[$\alpha$/Fe]$<$0.25, and [$\alpha$/Fe]$>$0.25 dex. The correction
to surface densities enhances the densities between 0.2 and 0.3 dex,
slightly shifted by $\sim$0.05 dex compared to the model, but this is
expected given that the scale heights from SEGUE were mapped on a
[Fe/H]-[$\alpha$/Fe], which is significantly different from the abundances
in APOGEE. However, the level of the high-$\alpha$ peak is  correct.

Figure \ref{fig:alphasfr}b and f shows the APOGEE data at the solar cylinder
(7 to 9 kpc from the Galactic centre) for stars with a signal-to-noise
ratio (S/N) greater than 150.  To ensure that frequency distributions were
not affected by the exact choice of S/N, we
checked that this only changes the relative significance of the peaks
by $<$10\%. Old populations at large distances from the Galactic plane
are much better sampled in these survey data than in our reference
sample (Fig. \ref{fig:alphasfr}a). The drop in and cessation of the
star formation activity for $\sim$1 Gyr leads to a strongly bimodal
distribution that is well reproduced by our model SFH. Figure~\ref{fig:alphasfr},
panels c and g and d and h, shows the distributions outside the solar vicinity
for the distance range 5 to 7 kpc and 3 to 5 kpc from the Galactic
centre. For this analysis, we adopted a less stringent S/N cut (only
80, compared to 150 used in the previous sample) to allow for a larger
number of stars and better statistical sampling of all populations,
especially the older, generally more distant stars. All plots illustrate
that SFHs 1, 2, and 3 are incompatible with the observed stellar
densities. None of these star formation histories is able to reproduce
the dip in stellar densities between 0.07$<$[$\alpha$/Fe]$<$0.17
dex. In contrast, SFH 4, with a sharp drop between 10 and 9 Gyr and
no or little star formation between 9 to 7.5 Gyr, agrees remarkably
well with both the local and the remote [$\alpha$/Fe] distributions,
reproducing both the peak centred on [$\alpha$/Fe]=0.02 dex and the sharp
decrease at [$\alpha$/Fe]$\sim$0.05 dex. SFH 4 explains the distribution
remarkably well considering it was not tuned in any way to fit these data,
but is precisely the SFH derived by fitting the evolution of [Si/Fe]
versus time using local data \citep{snaith14, snaith15}.  For SFH 2 and 3, the dip is not reproduced, but neither is the peak
of stars at 0.025 dex (thin disk) because of the effect mentioned in
the previous section: the number of SNIa produced early in the thick-disk phase in these models is small because the SFR during this period
is low and the iron ejected by SNIa is limited, which then affects the
decrease rate of [$\alpha$/Fe].

We conclude these comparisons by emphasizing the following important point:
the dearth of stars in the SFH between 7 and 8.5 Gyr was detected using
two entirely different approaches.  The SFH from Snaith et al.,
with a dip at $\sim$8Gyr, was derived from fitting the shape of the
relation of [$\alpha$/Fe] abundance $vs$ age alone, and in this case,
observed stellar densities played no role. As shown in \cite{snaith15}, the (abrupt) change
in the slope of [$\alpha$/Fe] abundance $vs$ age requires the dip in the
SFH to reproduce the [$\alpha$/Fe] abundance $vs$ age data.
Stellar densities from APOGEE now provide new and crucial evidence of a
rapid change in the star formation intensity between the two epochs of
thick- and thin-disk formation. We show below that the dip in the SFH has also been
found in local age distributions of stars (see Sect. \ref{sec:previousfindings}), although it has not
always been recognized as such.

\subsection{Effect of the IMF}

The IMF enters our comparisons in two ways. The first is through the
chemical evolution model, the second is through computing the number of
stars as a function of [$\alpha$/Fe]. We consider each of these in turn.
The IMF is an input of the model that was used in \cite{snaith15}
to adjust the observed age-[Si/Fe] relation by varying the SFH.  Hence,
changing the IMF affects the resulting best-fit SFH to the chemical
evolution of the MW.  It was shown in Snaith et al., however, that the
differences are small for a reasonable range of IMFs. This is illustrated
in Fig. 14a of Snaith et al., which shows that the SFHs obtained
for the Kroupa, Salpeter and Scalo IMFs all show a significant dip at
$\sim$8 Gyr. Only the IMF from \cite{baldry03} can reproduce the age-[Si/Fe]
relation without a dip in the resulting SFH. However, the model
generated with the IMF from \cite{baldry03} significantly overestimates the
[Si/Fe] abundances at all metallicities.  The best-fit model using the
Scalo IMF produces a dip in the SFH, but also significantly modifies the ratio
of thin to thick disks. Again, however, the model with a Scalo IMF results
in a significant offset of the inner-disk [Fe/H]-[$\alpha$/Fe] sequence by underestimating
[Si/Fe] by about 0.05 dex at a given metallicity.  We find that only
the Kroupa IMF enables us to fit all constraints given by elemental
abundances of stars within the solar vicinity.

\begin{figure}
\includegraphics[trim=40 670 0 20,clip,width=9.cm]{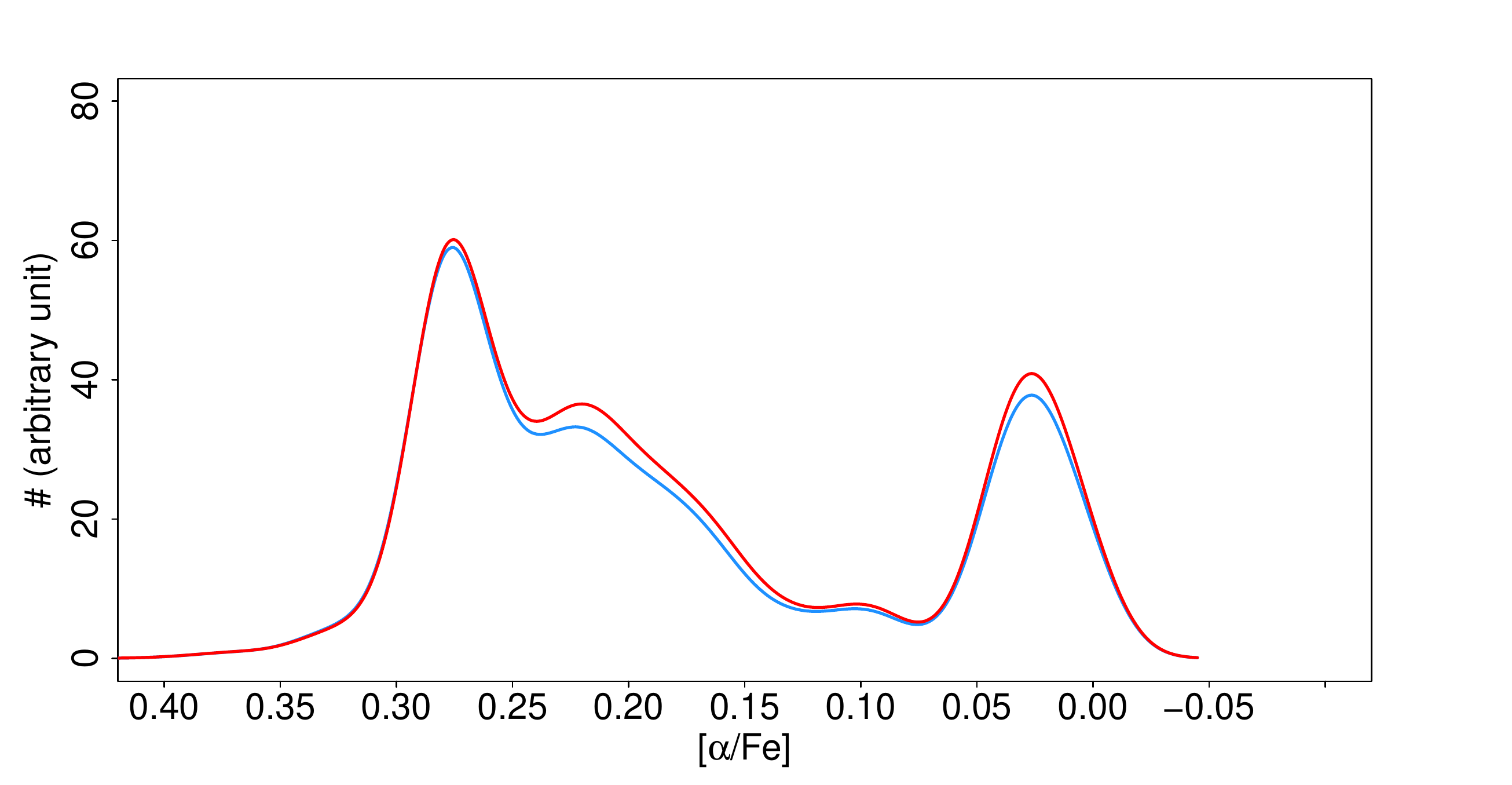}
\caption{Effect of the IMF on the relative mass distribution.The blue curve 
is derived using a \cite{kroupa01} IMF, the red curve with a Salpeter IMF.}
\label{fig:imfeffect}
\end{figure}

We now consider how the functional form of the IMF shapes the observed and modeled
stellar density distributions as a function of [$\alpha$/Fe].  The effect
of the IMF is likely to be significant if the range of masses that enters
the selected stars of Fig. \ref{fig:alphasfr} is large. In this case,
the IMF could have an effect on the ratio of the most massive to least
massive stars in the distribution.  However, this is unlikely to be the
case: the mass range of giants that enter the sample is rather small, limited
to about 0.4 M$_{\odot}$. Figure \ref{fig:imfeffect} shows that the effect is small, 
with the two curves obtained assuming a Kroupa (2001) and a Salpeter
IMF (blue and red curves).

\subsection{Radial migration}

What are the possible effects of radial migration on the present work?
Our analysis relies on a clearly identifiable inner-disk sequence in both the Adibekyan et al. data and the APOGEE survey. We recall that by inner disk, we mean both the thin- and the thick-
disks stars that make the clear -- and almost uncontaminated -- sequence visible in the APOGEE data at R$<$7 kpc
in the [Fe/H]-[$\alpha$/Fe] plane.
This sequence is also visible in the solar vicinity, but in continuity with the low-[$\alpha$/Fe] sequence. 
The outer disk (R$>$10 kpc) contains no or very little inner-disk stars and results from a different 
chemical evolution according to our description in \cite{snaith15}.
\cite{haywood13,haywood15} argued that the tight relations between age and metallicity 
(in the thick-disk phase) and 
$\alpha$-abundances for stars on the inner-disk sequence observed in the solar vicinity 
imply that the chemical evolution and star formation in the inner disk must have been remarkably
homogeneous and uniform. This is now confirmed by the similar chemical 
patterns seen in APOGEE at different Galactocentric distances (see Hayden et al., 2015, and Fig. 3). 
It may be that, as noted in \cite{halle15}, radial migration
has mixed the inner disk inside
the OLR, but mixing stars that have uniform properties, independent of Galactocentric distance, 
will not change any observed trend.

More generally, Fig. 4 of \cite{hayden15} shows that the inner-disk sequence clearly dominates at R$<$7kpc and the outer-disk sequence
at R$>$10kpc, with a limited overlap of the two  outside the 7-9 kpc ring and
almost none outside the 7-11 kpc ring: 
chemical patterns at R$<$7kpc are essentially different from
those at R$>$11 kpc. 
This confirms that the solar vicinity is a transition zone
between the inner and outer disk, as anticipated in \cite{haywood13}.
In this same article, we concluded that radial migration must have been very limited 
and cannot ``have redistributed stars in significant proportion
across the solar annulus'', which is what is observed in APOGEE.
These results favour models where the evolution of inner and outer disk  
are disconnected. By extension, they exclude models with extensive radial migration,  
where large portions of stars at a given radius come from large distances
(more than 2kpc) and where inner and outer disks have similar chemical patterns 
because they result from the same chemical evolutionary history \citep[see e.g.][]{minchev14}.
Hence, radial migration probably plays essentially no significant role in the distribution 
of $\alpha$-abundance analysed here.

\subsection{Possible biases}

Using the SEGUE survey, \cite{bovy12a} found that three effects
may conspire to produce an artificial bimodality in the star count
distribution of [$\alpha$/Fe] abundances: (1) the selection function of the
survey, (2) counting stars in volume rather than considering surface
densities, and (3) the binning of stars per unit [$\alpha$/Fe] instead
of mass per unit [$\alpha$/Fe], because lower metallicity stars have
lower masses, hence the metal-poor peak represents relatively less
mass than the metal-rich peak. According to \cite{bovy12a}, this last
effect is likely to be small because the difference in mass between
[Fe/H]=-1.5 and 0.3 dex is $\sim$20\%.  Figure \ref{fig:alphafehapogee}
shows that the two peaks cover an even smaller range, being separated
by $\sim$+0.5 dex, which means that the effect is likely to be significantly
smaller than even this estimate. 

We now consider points (1) and (2). \\
{\it (1) \textup{The selection function of the survey}.} \cite{hayden15} 
have shown that the sampling function of APOGEE does not introduce any
significant bias in the metallicity distribution functions (MDF). Although
there might be a bias in the [$\alpha$/Fe] distribution without producing
a bias in the [Fe/H] distribution, we note that the
MDF of APOGEE in the inner disk is clearly bimodal, with one peak at
+0.25 and another at -0.25 dex.  Since the MDF is not biased by the
selection function of APOGEE, we considered this as a real astrophysical feature.
Because of the good correlation of [$\alpha$/Fe] with metallicity within
the inner-disk sequence, it is expected that a bimodality in [Fe/H]
distribution directly corresponds to a bimodality in the [$\alpha$/Fe]
distribution.

We note that if the selection function of SEGUE were responsible for
producing the bimodality, we would not expect this feature to arise in
other surveys, but it is present in both APOGEE and GES \citep[see][]{recio14}.
We note also that in the case of solar vicinity samples, the
selection function of large surveys does not enter into consideration,
but the bimodality is still present.

\begin{figure}
\includegraphics[trim=10 220 0 220,clip,width=9.cm]{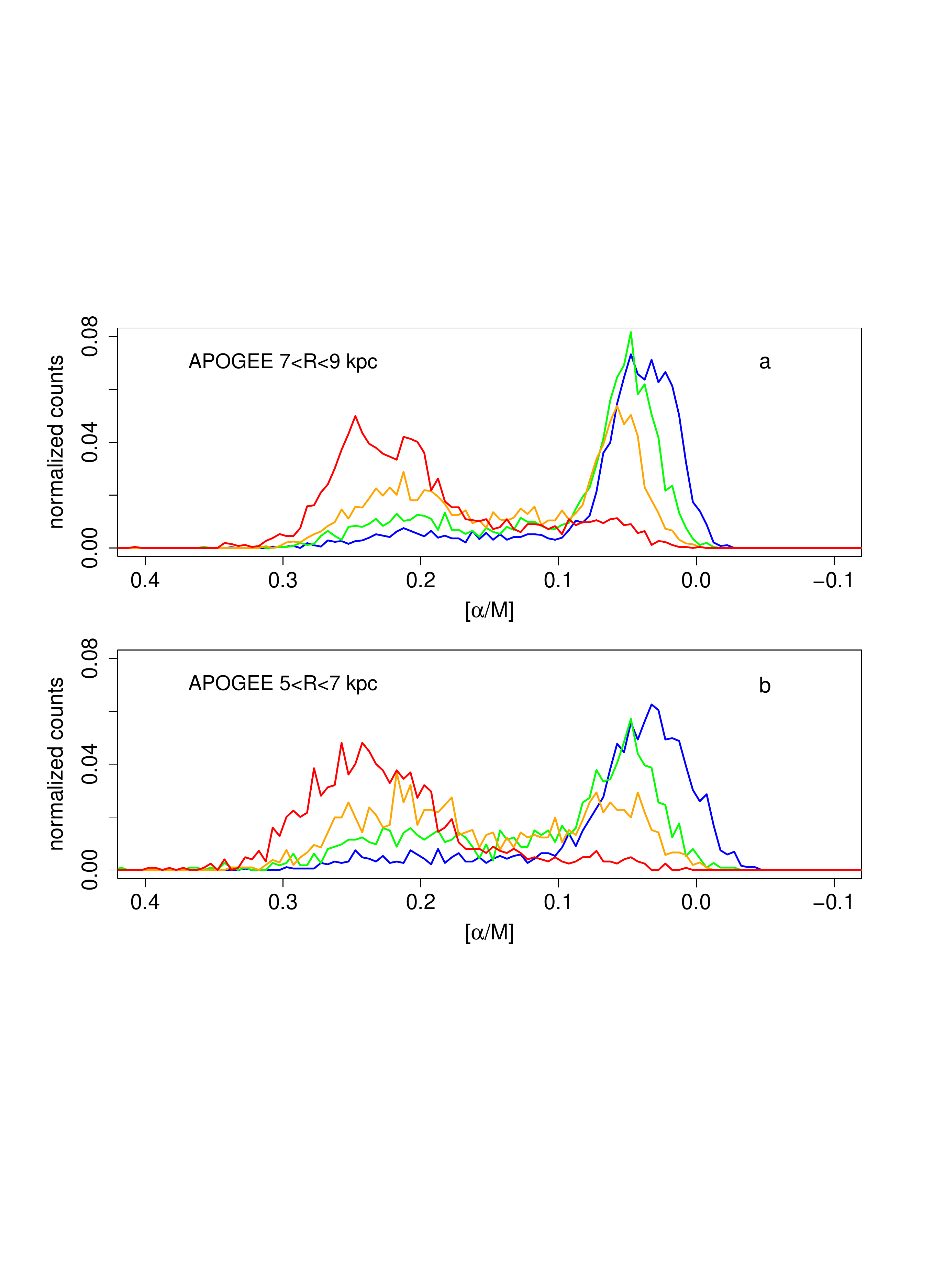}
\caption{ Normalized [$\alpha$/M] counts from APOGEE in two different
distance bins to the Galactic centre (the distances are indicated in each
plot).  The different curves correspond to different z intervals above
the Galactic plane. Red represents stars above 1 kpc, orange those between 1 and
0.5 kpc, green stars between 0.5 and 0.25 kpc, and blue stars
at 0.25 kpc relative to the
Galactic mid-plane.  At no height above the Galactic plane are the star
counts dominated by stars with [$\alpha$/M] between 0.08 and 0.18 dex.}
\label{fig:histalphaz}
\end{figure}

(2) It might be that this is an effect of the volume sampling (i.e. counting
stars in volume rather than considering surface densities). Converting volume densities into surface densities might alleviate
the bimodality.  Volume sampling of stars in the solar vicinity shows
the bimodality. If the conversion into surface densities were able to
alleviate the bimodality, it would imply that stars with intermediate
[$\alpha$/Fe] abundances (corresponding to the gap) are more numerous at larger
heights than they are close to the Galactic plane.  However,  at all
distances from the Galactic plane, the distribution of [$\alpha$/Fe] abundance
is either bimodal or dominated by stars that belong to either the
$\alpha$-poor or $\alpha$-rich peaks of the distribution. This is illustrated
very well in Fig.~\ref{fig:histalphaz}, which shows the normalized
counts of stars selected as explained in Sect. 3.1 (and which lie
above the black line in Fig.~\ref{fig:alphafehapogee}) as a function of
[$\alpha$/M] at different heights above the Galactic plane, 0-0.25,
0.25-0.5, 0.5-1.0, and above 1 kpc. It shows that
stars with [$\alpha$/M] in the range (0.08,0.18) dex are not
dominant at any height.  Hence,
there is no reason to expect that by simply summing the contributions
of stars at all disk heights we would fill the gap.

Beyond these considerations, some additional comments are
warranted. First, abundance determinations are more precise in APOGEE
than in SEGUE because of the higher spectral resolution and S/N of its data, which enhance the contrast between the peaks and the gap
of the distribution in APOGEE (cf., for instance Fig. 1 (left)
of \cite{bovy12a} and Fig. \ref{fig:alphasfr} shown here).  Second,
in estimating the [$\alpha$/Fe] abundance distribution, Bovy et al. selected
all stars in the ([Fe/H],[$\alpha$/Fe]) plane, including stars within
the metal-poor thin disk, which in the SEGUE survey do not appear as
a separate sequence because of the lower spectral resolution and poorer
determinations of the metallicities. The metal-poor thin-disk stars, however,
represent the separate chemical evolutionary history of the outer disk
(see Haywood et al. 2008, 2013, Snaith et al. 2015).  The oldest and
most metal-poor of these stars have [$\alpha$/Fe] abundances intermediate
between the $\alpha$-rich (old thick-disk) and $\alpha$-poor (thin-disk)
stars. Although such metal-poor outer-disk stars have similar $\alpha$
abundances, they have metallicities well below that of the thick disk at
the same age \citep{haywood13} and therefore would contribute to
filling in the gap in the stellar abundance distributions of the inner-disk stars. Hence, in attempting to understand the evolutionary history
of the inner disk, these stars should not be mixed into the analysis
(as we emphasized several times in this article).

\subsection{Previous findings}\label{sec:previousfindings}

The SFH derived in Snaith et al. (2014, 2015) is the first, and up to now,
the only, determination of the {\it \textup{general}} SFH of the MW. All previous
available studies were determinations of the local age distributions. It
is important to emphasize that age distributions determined from local
data {\it  \textup{are not}} star formation histories of the MW, although
they sometimes are inaccurately referred to as such. The contribution of thick-disk stars, in particular, are underestimated because while scale-height
corrections are sometimes applied to take the contribution
of older stars at larger distances from the Galactic plane into account, the thick
disk lies predominantly in the inner Galaxy and its contribution in the
solar vicinity is small.
The SFH derived from abundances assumes a model to fit the age-abundance
relation, but no density correction is needed. The derived SFH is general and representative
of the whole disk as far as the age-[$\alpha$/Fe] abundance relation is general 
and representative. 
If so, it means that the minimum we detected in the SFH in chemical data
in Snaith et al. (2015) and now in star count data (Sect. 3.3)  should also
be present in local age distributions.

\cite{cignoni06} determined an age distribution from the Hipparcos
HR diagram.  They found a distribution that shows a broad minimum
between ages of 6 to 10 Gyr, showing an upturn in the last bin between 10
and 12 Gyr.  Because the resolution in age that can be obtained from
inverting the HR diagram for old stars is poor and because thick-disk
stars are poorly represented in the solar vicinity, studies of this type
provide little information about the detail of the age distribution of
thick-disk stars.  In addition, as just mentioned, this population is
strongly underrepresented if volume corrections are not applied, as is
the case in this particular study.  However, because the corrections
are expected to be significant only for the oldest stars ($>$9-10 Gyr),
the dip found by \cite{cignoni06} would remain, in agreement with
our general findings.

\citet[see also Reddy et al., 2006]{gratton96} were probably the first to suggest from the analysis of 
chemical abundances the possibility of a gap in the SFH of the MW, at the transition of the thick to the thin disk.
\cite{bernkopf06} and \cite{fuhrmann11} suggested that an age gap of about 4 Gyr separates the thick and thin
disks, having found that the thick disk is a 12 Gyr old burst
population, with only few stars with intermediate ($\sim$10 Gyr) age. These findings
are roughly compatible with our own results, excluding the fact that the thick disk 
is a burst population, as already mentioned in the introduction.

\cite{rowell13} used the white dwarf luminosity function to investigate
the SFH in the past 9 Gyr.  He found that the SFH is bimodal with two
broad peaks at 2-3 Gyr and 7-9 Gyr, separated by {\it
`\textup{`a significant lull of magnitude 30-90\% depending on the choice of the
cooling models}''}.  Adopting the cooling models of \cite{fontaine01},
the estimated SFH shows a very significant pause extending from 5 to
7 Gyr with a minimum at $\sim$ 6 Gyr, compatible with zero star
formation activity.  As was also the case for \cite{cignoni06},
\cite{rowell13} determined an age distribution of stars in the solar
vicinity. Similarly, a true SFH would require volume corrections that
were not applied.  Here again, the most significant corrections would
apply to thick-disk stars.  Hence, if these corrections were
taken into account, the contrast between the SFR at ages $>$ 7 Gyr and
the SFR during the minimum would be increased. The lull detected by
Rowell would remain, and this lull again would generally agree
with our favoured SFH.  Another significant difference is the fact that
the minimum detected by \cite{rowell13} occurs at 6 Gyr, while in our case
the SFH is minimum at $\sim$8 Gyr.  This is not as critical as it seems,
however, because there are still large differences between isochrone
datings and ages deduced from white dwarfs.  For instance, \cite{fontaine01}  
estimated the age of the disk to be of the order of 11 Gyr,
or at least 2 Gyr younger than the age we determined for the oldest
thick-disk stars \citep{haywood13}.  Kalirai (2012, see also Kilic
et al. 2010) measured an age for the white dwarfs from the Galactic halo
of 11.4$\pm$ 0.7 Gyr, significantly younger than local halo subdwarfs
(see for example \cite{vandenberg14}, who derived ages of 12$\pm$0.14,
12.56$\pm$0.46, and 14.27$\pm$0.38 for three local subgiants). Similarly,
ages of globular clusters derived from white dwarfs can also be younger
than those obtained from isochrone fitting technics: for instance,
\cite{hansen07} derived an age for NGC 6397 of 11.47$\pm$0.47 Gyr,
while \cite{gratton03} for instance derived an age of 13.9$\pm$1.1 Gyr
\citep[see comments in][]{torres15}. 

The conclusion from these comparisons is that using three entirely different
datasets and methods, three studies found evidence of a significant dip
in the age distribution of stars in the solar vicinity, supporting our
general findings.

\section{Discussion}\label{sec:discussion}

The data from APOGEE shown in the previous section include stars
from 3 to 9 kpc from the Galactic centre and demonstrate that the dip
found in the SFH is a pervasive feature that occurs throughout the
disk. This implies, unlike any other previous result or analysis, that
the cessation of the star formation activity was a general phenomenon
that did not only occur at the solar cylinder. Remarkably, in less than
2 Gyr, from ~10 to 8.5 Gyr ago, the star formation rate has decreased by
more than a factor of 10. At a redshift of around 1, the MW was
on its way to become a lenticular galaxy (see also Sil'chenko 2013), 
and then the star formation activity resumed (see Sect. 4.4 for a possible
explanation).
We now discuss the consequences of this finding.

\subsection{Thick and thin disks: are they the same population?}

Recently, \cite{bovy12a,bovy12b} has discussed the possible continuity
between thick- and thin-disk stars.  Their arguments were twofold. They first
concluded that the dip in the [$\alpha$/Fe] distribution is only apparent and not
due to astrophysical causes, but is instead due to selection and volume
effects. This conclusion is not confirmed by our analysis, as we explained
in Sect.  3.2.  The second argument is based on the measurement of
structural parameters of mono-abundance populations from star counts,
which shows a smooth transition in scale height and scale length from thick- to thin-disk stars. This analysis has been confirmed from APOGEE data
by \cite{bovy15}. We return to this argument below,
but we first wish to add two important facts that must be taken into
account in this debate.  It was suggested three decades ago that the
thick disk pre-enriched the thin disk \citep{gilmore86}.  Indeed,
although there is a break in the slope of the age-[$\alpha$/Fe] relation
that clearly points to two separate epochs of star formation, there is
strict continuity in the chemical properties of the two populations (as
shown in Fig. 6 of \cite{haywood13} and Fig. 3 of \cite{haywood15}):
 the thin inner disk originally inherited the chemical traits of the thick disk.
The change of slope in the age-[$\alpha$/Fe] relation is due to a
change of the intensity of the star formation between the two epochs.
The present findings now confirm that the transition between the two disks
probably occurred during a period of almost complete cessation of the star
formation activity.  Hence, although there is continuity in elemental
abundances and possibly in structural properties, the two populations
are well differentiated in their star formation histories: the thick
and thin disks formed their stars with two distinct phases of Galactic evolution.

How can we account for the results showing a continuity in the structural
parameters under these circumstances?  The thick-disk population
cannot be described with a single scale height. The analysis of \cite{bovy12b, bovy15}
showed that the $\alpha$-rich population can be decomposed into a
series of subpopulations whose scale height decreases with increasing
metallicity. This is confirmed by data from stars in the solar vicinity,
where it has been found that the vertical velocity dispersion continuously
decreases with age within the thick disk (Haywood et al. 2013,
Fig. 11).  This means that although the gap in time between the start
of one population and the end of the other was relatively large, 
any discontinuity left in the structural parameters of these
two populations is probably small simply because the structural properties 
of the young thick disk and the old thin disk are very similar.
In addition, various dynamical effects may have blurred possible
differences in kinematics and spatial distributions of the populations
of the oldest thin-disk and the youngest thick-disk stars over
the past 8 Gyr.

In view of these results, the discussion of a continuity (or not) between
the two populations, based on a decomposition of structural parameters of
``mono-abundance subpopulation'' appears contrived.  The use of structural
parameters as defining properties of a stellar population is implicit in
the definition of all MW stellar populations, which are often viewed as
spatial entities (``disk'', ``bulge'', ``halo'').  But can we define
a continuity between two populations solely on the basis of their
structural parameters?  The MW has a ``bulge'' , but possibly no separate
bulge population of its own. Recent studies such as those by \cite{shen10}, 
\cite{kunder12}, and \cite{dimatteo14, dimatteo15} suggest that the bulge
is basically a structure formed from disk stars through dynamical disk
instabilities, a ``pseudo-bulge''.  Hence, defining a population from its
structural parameters is ambiguous at best.  More accurately, a stellar
population ought to be defined by parameters that are linked
as directly as
possible to the ISM out of which its stars formed and at the time
they were formed.  From this point of view, the thick and thin
disks are clearly separate but parent or child populations.

\subsection{Star formation quenching in the Milky Way $vs$ other galaxies}

We interpret the decrease in the star formation activity in the MW
10 Gyr ago (z$\sim$2) and the lull in the SFR at 7-8.5 Gyr as a
manifestation of the quenching.  Quenching is observed in distant
galaxies and has several defining characteristics.  First, quenching
in the MW is extended in time and represents a significant decline in
the SFR by more than an order of magnitude in less than a Gyr, and
this insignificant level of SF is maintained for about 1 Gyr.  This agrees well with the observed quenching of MW-type galaxies in
the early Universe \citep{morishita15}.  Morishita et al. selected
Milky Way-type progenitors at different redshifts using the technique of
abundance matching where they studied the redshift evolution
of star-forming
galaxies with co-moving densities similar to MW-type galaxies in the local
volume.  They observed that the inner parts of the galaxies
they selected statistically become redder at z$\sim$1.6 and then quench their star
formation at z$\sim$1.0, in very good agreement with our time-line for the
star-formation rates of the MW.  We emphasize that a difference must be
made here between {\it \textup{quenching}} of the star formation activity and the
transition from an active to a {\it \textup{quiescent}} phase of star formation.
\cite{schawinski14} for instance argued that quenching in late-type
spirals is the slow transition from an actively star-forming galaxy
to one that is quiescent as a result of the exhaustion of its gas.  Clearly,
the quenching observed in the inner disk of the MW is not a slow transition to
quiescence.  The drop in the SF activity in the MW was fast --
less than 1 Gyr, but could be significantly more rapid than that since
this is below the time resolution of our SFH. This means that while the transition
to quiescence in the MW was fast, just as it appears to be in early-type
massive galaxies, the MW is in strong contrast to early-type galaxies
because its quenching phase was followed by the re-initiation of its star
formation. This re-initiation of star formation lasted until today,
formed roughly 50\% of the MW's stellar mass, and the star-formation rate
was roughly constant.  Between 9 and 7 Gyr, it is possible that the
only significant SF activity occurred in the outer parts of the disk
(R$>$10~kpc, see Haywood et al. 2013). However, because of the low
gas and stellar densities in these outer regions, the star formation
efficiency was likely very low and the integrated contribution of such
star formation in the outer disk must have been small.

Phenomenologically and in particular, quenching in galaxies is observed
through their bimodal distribution in colour-magnitude planes, whereby
the quiescent, quenched galaxies form a red sequence, while the still
actively star-forming galaxies form a blue cloud (see e.g. Gabor et
al. 2010 for a nice summary).  The present position of the MW,
whether in the blue cloud, in the green valley, or in between them,
is unknown, essentially because of the uncertainties that plague the
colour measurements of our Galaxy (Mutch et al. 2011, or more recently
Licquia et al. 2015). It is possible, however, to comment on the possible track between the blue and
red sequences of the MW at the time of its quenching. During the
period of the thick-disk formation, the Galaxy was actively forming stars
and therefore belonged to the blue cloud. The duration of the passage
from the blue cloud to the red sequence depends on the quenching
rate experienced by a galaxy.  The faster the quenching, the
faster a galaxy crosses the green valley.  For a decrease of the SFR
by an order of magnitude in 2 Gyr, as indicated by Fig.~2, the quenching
timescale is $\sim$0.86 Gyr, which would correspond to a crossing time
of $\sim$1Gyr \citep{martin07}. This means that at the end of the
quenching phase, the MW was at the limit of the red sequence.
Similar estimates from \cite{blanton06} are that a sharp cut-off in star
formation induces a migration to the red sequence in $\sim$1 Gyr. The
SFH of the MW is consistent with this.

The second reason to consider the observed lull in the SFH as a
manifestation of quenching is that it is associated with a morphological
transformation,  marking the transition from the thick to the thin disk.
Hence, it is not any fluctuation of the SFH, but one that corresponds
to a major phase in the history of the Galaxy, clearly reminiscent of
the transitional phase that is detected in galaxies at high redshifts.
A crucial difference to the standard interpretation is, however, that the
quenching in the MW is linked to the thick disk, not to the bulge.
The MW is most probably a pure disk galaxy \citep[see e.g.][]{shen10, dimatteo14, dimatteo15}.
That is, its classical bulge,
if it exists, is small (B/T$<$0.1).  In contrast, its thick disk
is massive, totaling about half the stellar mass of the entire disk.
We argue in the next section that bulge quenching, observed
to take place in high-redshift samples, may often be thick-disk quenching instead.

We note that the quenching episode is also contemporaneous with other events
in the MW.  According to \cite{haywood13}, the outer disk
ignited star formation at $\sim$ 10 Gyr. The time at which  the SF
starts to slow down in the inner disk therefore coincides with the time when star
formation activity ignites in the outer disk.  The minimum of the star
formation (at z$\sim$1) may also be contemporaneous with the birth of
the bar.  Although the precise time of the bar formation is not known,
galaxies of the mass of the MW are thought to form their bar at z$\sim$1
\citep{sheth08, melvin14}.  Interestingly, this is
also the epoch at which \cite{morishita15} detected an increase in
stellar mass in the central regions of MW progenitors, at a time when
their galaxies have already quenched their star formation activity.
Bars have been proposed as a possible agent of quenching in late-type
galaxies \citep[e.g.][]{gavazzi15}.

Another interesting feature from \cite{morishita15} is that they
clearly observed that quenching occurs in the inner regions, while apparently
SF continues in the outer parts. This is also compatible with the
sequence of events described here for the MW, when the star
formation activity shuts down in the inner regions while it ignites in
the outer disk.

\subsection{Are these results compatible with the general history of
late-type galaxies?}

The quenching episode measured here shows that galaxies like the MW
may have had large variations of their SFR in their past, with a decrease
by an order of magnitude occurring on a timescale of $\sim$1 Gyr.  The question is whether this variation is consistent with the observed dispersion of
the main sequence of galaxies.  Comparisons of our SFH  with the main
sequence have been presented in \cite{lehnert14}, who
showed that the MW lies systematically below the star formation
rate that would be expected for a typical galaxy of its mass.  Figure 2 of \cite{whitaker15}, 
for example, shows all galaxies plotted in the main-sequence plane (either active or quiescent).
It is evident there that with a star
formation rate of a few solar masses per year and because of its
own mass, the MW has been a quiescent,
or almost quiescent, galaxy since z$<$1. This is compatible with its position in the green valley
\citep[see][]{mutch11, licquia15}.

In the thick-disk formation phase, our Galaxy had its most intense star
formation activity, but also lay below the main sequence.  We
note that the MW is not an exception in this regard: for MW
progenitors, \cite{vandokkum13} measured a star formation
activity similar with what we measured for our Galaxy.

Assuming a total current stellar mass of 5$\times$10$^{10}M_{\odot}$, the MW 
sustained an SFR of between 10 and 15  M$_{\odot}$.yr$^{-1}$ during the
thick-disk formation (see Fig. 1), and 10 Gyr ago (z$\sim$2), the MW
was half way to building its thick disk (with a stellar mass of the order of
1 to 1.5 $\times$ 10$^{10}$M$_{\odot}$).  For a
galaxy of this mass,  the main sequence at redshift 2 is nearer to 30-40M$_{\odot}$ yr$^{-1}$.  However,
at a rate of 40M$_{\odot}$ yr$^{-1}$, the MW would have formed all its
mass within 1.3 Gyr, which implies that it cannot have sustained this star
formation activity for very long.  A bursty mode that, while keeping a mean
SFR of 10-15 M$_{\odot}$.yr$^{-1}$, alternates  short phases of intense
star formation  with more quiescent periods is not favoured, however,
because it would probably inject sporadically high levels of metals in
the ISM and would produce a dispersion in abundances that is not observed
(of course unless the metal loss was tuned to keep the dispersion in the
metallicities of the stars low), with the [Fe/H]-[$\alpha$/Fe], the age-[$\alpha$/Fe] 
and age-metallicity relations all being  tight during the thick-disk phase
\citep{haywood13}. 
This low dispersion of abundances and metallicities 
of stars at constant age is most simply explained by the MW having kept
an approximately constant rate of star formation of 10-15 M$_{\odot}$.yr$^{-1}$
during the phase of thick-disk growth. All this implies that our Galaxy,
and possibly most galaxies of its type, has remained quiescent
(in a relative way, this largely depends on the definition of the
main sequence) for most of its history, possibly not lying along the
ridge line that defines the main sequence of galaxies.

\subsection{What is the origin of this relatively brief quenched phase?}

In the particular case of the MW, it is difficult to associate
the shutting down of star formation and its reactivation with a shortage
of gas and then significant accretion of fresh fuel, since for such a scenario
we would expect to see a corresponding signature in the evolution of
the chemical abundances.  In contrast, the age-[$\alpha$/Fe] relation shows
a clear continuity (see Fig. 6 of \cite{haywood13} for example),
which is very well represented by a closed-box model \citep{snaith15}.
We therefore exclude the possibility that the quenching episode in the MW
is directly tied to its history of the gas accretion.

Morphological quenching has been proposed by \cite{martig09}
for galaxies with substantial spheroidal components, or thick disks,
which could stop star formation by stabilizing the gas disk.  At the
end of the thick-disk phase, before the thin inner disk started to form, most of the MW stellar mass was located in a component which, with
a scale length of 2~kpc \citep{bensby11, cheng12, bovy12b}, was strongly concentrated in the inner regions. 
Although \cite{martig09} clearly envisaged that morphological quenching could apply
to thick disks, we emphasized in Sect. 2.1 that at the end of the thick-disk phase, the gas fraction in the MW was still likely very high
($\sim$50\%). This is far from the values of the case studied by \cite{martig09}, where the gas fraction stays essentially below 15\%
throughout the transformation from disk to spheroid-dominated galaxy.
Hence, it is not clear how efficient morphological
quenching would really be in the \textcolor[rgb]{1,0.501961,0}{\textcolor[rgb]{0,0,0}{\textcolor[rgb]{0,0,0}{MW}}}{
}.

Other possibilities for quenching have been proposed more recently. For
instance, it has been suggested that the heating of the gas from the radiation field of
low-mass stars or the winds of dying low-mass stars
could prevent star formation in galaxies, see \cite{kajisawa15}
and \cite{conroy15}.  We also noted in the previous section that
the formation of the bar in the MW is possibly contemporaneous
to the quenching phase. \cite{athanassoula13} and \cite{gavazzi15}
proposed that the bar might be responsible for the quenching of
the star formation by sweeping the gas within the corotation radius to
the central regions through the loss of angular momentum.  While this
is an attractive possibility, it is not clear in this scenario how to
ensure that the star formation activity resumes to form the
disk stars and -- if it is through the accretion of fresh gas -- how to
ensure the chemical continuity between the thick- and thin-disk stars.
Moreover, in the scenario proposed by \cite{gavazzi15}, the fraction
of gas assumed is 5\% of the stellar mass, which is far from the case 
being considered here, where half the mass of the disk could still be in gas. 
While in the scenario of Gavazzi et al. the gas is driven into the Galactic centre
and triggers a starburst, it is difficult to imagine here that 
the same would occur in our case because  the  quantities of gas involved are so much larger
and its conversion into stars would leave remarkable signatures.

Associating the shutdown in the star formation of the MW with both the end
of the thick-disk building phase and a bar is a very attractive solution
\citep[e.g][]{masters12,cheung13}.
However, as we noted, the bar cannot do this by redistributing the gas
permanently because the star formation needs to be re-ignited after $\sim$1-2
Gyr. Instead, we propose that the bar simply inhibits star formation
by shearing the gaseous disk, which through viscosity is transferred into
turbulence \citep{fleck81, renaud15}. The high level of turbulence
then provides pressure support for the gas in the disk, preventing it from
being swept into the circum-nuclear region of the MW \citep{athanassoula13}.

To investigate this possibility, we adopted a simple two-phase disk model.
One disk is purely stellar with an exponential scale length of 1.8 kpc, the other is purely gas with a scale length of 3.8 kpc \citep{bovy12b}.
Both disks are assumed to have equal masses, a total mass
of 5 $\times$ 10$^{10}$ M$_{\sun}$, which is approximately the current
stellar mass of the MW, the rotation speeds of the disks are 170 km
s$^{-1}$ and 220 km s$^{-1}$, and represent the thick and thin disks,
respectively. If both disks have constant stellar velocity dispersion,
we find from investigating the two-fluid stability of such a disk \citep[see][]{elmegreen95} that the disk is Toomre stable, that is, it has Q$\sim$1 for
a gas dispersion of $\sim$50 km s$^{-1}$ or higher everywhere in the
disk. The disk is most unstable to two-fluid instabilities at a radius
of about 3 kpc, and this region requires the highest energy
input to stabilize the disk.  Changing the assumptions in this model
within the observational constraints available, such as increasing the
disk scale-lengths (for instance, to 2 and 5 kpc for the stellar and gas disk,
respectively), little changes the exact value of the velocity dispersion
needed to stabilize the gas disk (i.e. Q$\ga$1).

The flat rotation curve of the gas disk with a velocity of 220 km s$^{-1}$
at 3 kpc has a local shear rate of about 80 km s$^{-1}$ kpc$^{-1}$.
If we assume that the bar has a co-rotation radius of
about 5 kpc 
\textcolor[rgb]{0,0,0}{\textcolor[rgb]{1,0.501961,0}{\textcolor[rgb]{0,0,0}{(see \cite{gerhard11} for a discussion about the uncertainty in the co-rotation radius, 
with reported values between 3.4 and 7 kpc;}} {}} \cite{wegg15} found 
that co-rotation must be beyond 5$\pm$0.2 kpc.), then the shear rate at 3 kpc is $\approx$$\Delta$V/$\Delta$r,
where $\Delta$V is the velocity difference between the bar and gaseous
disk, which is approximately 100 km s$^{-1}$ kpc$^{-1}$.  The shear
generated by the bar is stronger than or equal to that within the gaseous
disk itself.  If we assumed a larger co-rotation radius, the contribution
of the bar to the shearing rate would be proportionally higher (it increases
linearly with radius). If we assume that the gaseous disk has a thickness
of about 300 pc (which is approximately the scale height of the old
stars in the thin disk today), we can estimate the velocity dispersion
of the clouds by equating the energy of the shear to that dissipated by
turbulence as the shear rate times the disk thickness \citep[see][]{fleck81},
yielding 180 km s$^{-1}$ kpc$^{-1}$ * 0.3 kpc $\approx$50 km s$^{-1}$.
Of course, a key parameter to determine whether this idea is energetically
feasible is the dissipation time of the turbulence generated by the shear
from the gas disk rotation and the bar. The cloud-cloud collision time
is $\approx$50-100 Myr (the cloud-cloud dispersion within the ensemble
of clouds will decay on several cloud-cloud collision timescales;
e.g. \cite{silk09}).  The rotation time of the bar is about 180
Myr (compatible with the wide range of the measured values of the present-day bar pattern speed, \cite{gerhard11}),
while the stirring from the shear in the flat rotation curve of the
gas disk is injected constantly. Given the similarity of the timescales
and the long cloud-cloud collision time, it is plausible that the bar
plus disk can generate sufficient shear over approximately the necessary
timescales to maintain the turbulence that makes the disk globally stable.
To sustain the high rate of shear for the duration of the pause (taking into account
the decreasing phase of the SFR, this is from about 10 to 7 Gyr) requires
the bar to exist for about 15 rotation periods, which is plausible.

Of course, our arguments are qualitative and our discussion was only
intended to investigate the plausibility of such a scenario. The scenario
has some parameters in it for which we used the best observational
constraints available (in the rotation speeds, scale lengths, disk
thicknesses, velocity dispersions, etc).  We are sensitive to a few
of them.  For example, if the disk is thinner than we have assumed,
$<$0.3 kpc, the turbulence velocities will be lower, as would the shear be if
we assumed a shorter bar.  However, highly turbulent velocities would naturally thicken the gas disk, and 0.3 kpc is approximately that of the old
thin-disk stars \citep{bovy12b}.  The co-rotation radius of the bar
must be larger than 3 kpc or it will not enhance the turbulence where Q reaches its minimum, and it must be sufficiently long to inhibit
star formation out to large radii, although the gas disk is stable out
to large radii with only a modest enhancement in the turbulent velocity
(20 km s$^{-1}$), and so this is a weak constraint. While in detail we
are sensitive to the exact numbers we have chosen, they are plausible
given the observations and are not such that they must have particular
values for our scenario to be plausible.

We envision a transfer of energy from the large-scale shear into
the random cloud motions and then into turbulence in the clouds
themselves. While a detailed discussion of this is beyond the scope
of this paper, the transfer of energy may prevent the gaseous disk
from being gravitationally unstable on any scale (from the disk down to
individual clouds, \cite{guillard15}).  If the bar weakens over time,
then its effectiveness of shearing the gas will diminish, allowing the
thin disk to begin forming stars.  As noted in \cite{haywood13},
it may be that the thin disk started forming in its outer regions,
which would be a natural outcome of using the bar to stir the inner-disk gas, maintaining its stability.  We also note that a bar in the
thick disk will not be similar to the bars observed or modeled locally.
Bars in local disks are composed of thin-disk
stars and co-rotate with both the gas and stars at this (co-rotation)
radius.  At co-rotation, gas piles up as a result of orbital crowding at the
end of the bar, the piled-up gas fragments, eventually leading to star
formation \citep{renaud15}.  In the scenario proposed here, the bar still has a orbital velocity even at its ends that is much lower than
the gas, which will then not lead to orbital crowding, but will
generate significant shear.  In addition, again, unlike simulations of
bars consistent with local galaxies, the turbulence generated by a bar
as proposed here is significant compared to the rotational support of
the gas disk (50 km s$^{-1}$ compared to 10-20 km s$^{-1}$; Renaud et
al. 2015).  Thus, instead of the gas falling deeper into the potential
well after the development of a bar, the turbulent pressure partially
compensates for the loss of rotational support in a high-dispersion
system as proposed here (e.g. Burkert et al. 2010).  This overcomes
our previous objection that the bar will sweep gas into the centre when
quenching the galaxy \citep{athanassoula13,gavazzi15},
which would be inconsistent with the chemical continuity
between the stars of the young thick and old thin disks.  In our scenario,
star formation is prevented by turbulent support, which similarly supports
the gas from generally being swept into the Galactic centre through loss
of rotational support and dissipation.

\subsection{Mass-quenching in MW-type galaxies is thick-disk building}

\cite{vandokkum13} and \cite{morishita15} found consistently
that MW-type galaxies build their disks in a self-similar way at z$>$1.
This is what we would expect if these objects were building their thick
disks, growing in mass while keeping their scale length at a constant
value.  For the MW, it has been argued in \cite{haywood13,haywood15} that the thick disk grew in mass while keeping its scale length
constant at $\sim$1.8~kpc from an age greater than 12 Gyr to $\sim$9 Gyr
(or z$>$2 to z=1.6).  That is, the MW disk did not form inside-out,
but in a self-similar way, as found by \cite{vandokkum13} and \cite{morishita15}  
for MW progenitors.  These results strongly
indicate that the first substantial component formed in these objects
were thick disks, and therefore that quenching is related to thick-disk
building and not to the build-up of classical bulges
in MW-type galaxies.  This is also expected from statistics of bulges and
thick disks in the local Universe because of the ubiquity of thick disks in
galaxies \citep{comeron11} and the rarity of true classical bulges
in local galaxies: the local volume ($<$ 11 Mpc) is dominated by pure
disk galaxies  (Kormendy et al. 2010, Fisher \& Drory, 2010, see also
Laurikainen et al. 2014).  Recent results,
shown in \cite{dimatteo14, dimatteo15}, for example, have confirmed the view
that the MW is a pure disk galaxy, adding consistency
to the overall picture.  This statement is therefore likely to be valid
more generally for late-type galaxies, first because classical bulges
are rare among them, but also because thick disks are observed to be
relatively more massive in lower-mass disk galaxies \citep{comeron11}.

In contrast, more massive galaxies seem to grow inside-out \citep{vandokkum10, morishita15},
with massive bulges already in place at z$\sim$2.4  \citep{morishita15}.  This is what we would expect if
these objects first formed a bulge, then built their disk.
We note that for disks galaxies with stellar masses higher than that of
the MW, classical bulges are a more substantial fraction of their
total stellar masses.  S0-Sb galaxies, which tend to be more massive,
usually have high bulge-to-total-mass ratios. These massive bulges and
high bulge-to-disk-mass ratios are difficult to explain through secular
growth processes alone \cite[see][]{kormendy04}. These galaxies
mainly contain a classical bulge in their central regions and are most likely
the descendants of at least a fraction of the massive galaxies that are
observed by \cite{vandokkum10} and \cite{morishita15}. This
has to be the case because they generally selected their galaxy sample
through abundance matching (selected to have densities lower than a
constant co-moving density threshold independent of redshift), and those
samples contain galaxies that will no doubt be substantially more massive
than the MW at z=0.

If the phenomenon of quenching is related to the build-up of the
central parts of galaxies, it is therefore tempting to suggest that
thick disks may be more generally related to quenching for galaxies
on the low-mass side ($\la M^*$) of the stellar density distribution
of galaxies.  In contrast, spirals similar to M31 or more massive
usually contain a significant classical bulge (B/T=0.31 for M31, see
Kormendy et al, 2010), but have relatively low co-moving space densities.
Quenching could be linked to classical bulges in larger spirals and
more generally in massive galaxies with large B/T, or galaxies on the
high-mass side of the galaxy stellar mass distribution ($>M^*$).

\section{Conclusions}

Because [$\alpha$/Fe] abundances are tightly correlated with age, their number distribution
is an accurate record of the past star formation intensity in our Galaxy.
\cite{snaith14,snaith15} used the fact that the age-[$\alpha$/Fe] relation contains a fossil
record of the past star formation intensity
to derive a star formation history of the MW inner disk from a local sample of stars.
In the present study, we showed that the extensive abundance data from the recent APOGEE survey
confirm our previous study and in particular provide strong evidence of a lull in the 
SFH, similar to the phenomenon of quenching of high-redshift galaxies.

Quenching is a phenomenon that may be a generic phase in the life of
galaxies, and we found that the MW offers the best opportunity to study
quenching in significant detail.  To our knowledge, this is the first time
that a specific episode in the general evolutionary history of galaxies
is this directly confirmed from the archaeological record embedded
in the stars of our own Galaxy.  $\alpha$-elemental abundances show that a
significant drop, by a factor of 10, and long pause in the star formation
activity occurred in the MW between the age of 9.5 Gyr (z$\sim$1.6)
and 7 Gyr (z=0.8), with an almost complete quenching at z$\sim$1.0. This
event manifests itself in the lower density of stars along the inner-disk
$\alpha$-elemental sequence between $\sim$0.07 and 0.17 dex. This dip in the
[$\alpha$/Fe] abundance distribution, which occurs along the inner-disk sequence,
separating the thick disk and the inner thin disk, has been attributed to the combined
effect of the selection function and volume sampling of the stars. We
argued here that this dip is a real astrophysical phenomenon and that the
corresponding lull in the SFH is detected in several independent studies.

This quenching episode marks the end of the thick-disk formation
and is followed by a more quiescent phase of star formation in our
Galaxy. Although this event marks a clear discontinuity between the
formation of the thick and thin disks, we emphasize that there is
elemental abundance continuity between the two populations, so
that the end
of one phase is related to the beginning of the next despite the long
pause in the MW's star formation.  Hence, the origin of quenching does
not appear to be related to a shortage of gas in our Galaxy, otherwise
the necessary replenishment of the gas to re-ignite star formation would
have left a discontinuity between the abundance properties of the thick
and thin disks that we do not observe.

Although we were unable to associate this quenching episode with any of
the processes that have been proposed for the quenching in disks, we
emphasize that quenching in the MW is not the gradual transition
to quiescence, but seems to operate in a similar way and over
a similar timescale as in early-type
galaxies.  We proposed a scenario that
might link the cessation of star formation to the development of a strong
bar in the thick disk, and its subsequent weakening allowed star formation
to re-ignite.

On the basis of the close similarity between the evolution of the MW
and MW progenitors and from the demographics of thick disk
and classical bulges in the local Universe, we argue that the quenching of the star formation activity for late-type
galaxies is associated with the 
cessation of the growth of thick disks and perhaps with the start of the
growth of thin disks.

\begin{acknowledgements}
We are grateful to the referee for a detailed report and helpful comments.
MDL would like to thank Pierre Guillard for interesting and very helpful
discussions on shear and turbulence in galaxy disks.
\end{acknowledgements}

\clearpage

\end{document}